\documentclass[11pt]{article}

\usepackage{authblk}
\usepackage[english]{babel}
\usepackage[latin1]{inputenc}
\usepackage{lmodern}
\usepackage[T1]{fontenc}

\usepackage{amssymb, amsmath, amsthm, blkarray}
\usepackage{graphicx, tikz}
\usepackage{natbib}

\definecolor{mylightyellow}{rgb}{1,1,.8}
\definecolor{mylightgreen}{rgb}{.8,1,.8}
\definecolor{mydarkred}{RGB}{178,34,34}
\definecolor{mydarkgreen}{RGB}{34,139,34}
\definecolor{mydarkblue}{RGB}{72,61,139}
\definecolor{mydarkyellow}{RGB}{218,165,32}

\usepackage{marginnote}
\usepackage{soul}

\usepackage[a4paper,margin=30mm]{geometry}

\usepackage[]{hyperref}
 \hypersetup{
 colorlinks=true,breaklinks=true,
 urlcolor= mydarkblue,linkcolor= mydarkblue,citecolor= mydarkgreen,
 pdfauthor={Brigo, D., Huang, X., Pallavicini, A., S\'aez de Oc\'ariz Borde, H.}
}

\theoremstyle{plain}
\theoremstyle{definition}

\usepackage{eurosym}

\DeclareMathOperator*{\argmin}{arg\,min}

\title{Interpretability in deep learning for finance: \\ a case study for the Heston model}

\author[1]{Damiano Brigo\thanks{\tt damiano.brigo@imperial.ac.uk}}
\author[1]{Xiaoshan Huang\thanks{\tt xiaoshan.huang19@imperial.ac.uk}}
\author[1,3]{Andrea Pallavicini\thanks{\tt a.pallavicini@imperial.ac.uk}}
\author[2]{Haitz S\'aez de Oc\'ariz Borde\thanks{\tt haitz.saez-de-ocariz-borde17@imperial.ac.uk}}

\affil[1]{\small Department of Mathematics, Imperial College, London SW7 2AZ, United Kingdom.}
\affil[2]{\small Department of Aeronautics, Imperial College, London SW7 2AZ, United Kingdom.}
\affil[3]{\small Financial Engineering, Intesa SanPaolo, largo Mattioli 3 Milano 20121, Italy.}

\date{
\small First Version: December 1, 2020.  This version: \today
}

\begin{document}

\maketitle

\begin{abstract}

Deep learning is a powerful tool whose applications in quantitative finance are growing every day. Yet, artificial neural networks behave as black boxes and this hinders validation and accountability processes. Being able to interpret the inner functioning and the input-output relationship of these networks has become key for the acceptance of such tools. In this paper we focus on the calibration process of a stochastic volatility model, a subject recently tackled by deep learning algorithms. We analyze the Heston model in particular, as this model's properties are well known, resulting in an ideal benchmark case. We investigate the capability of local strategies and global strategies coming from cooperative game theory to explain the trained neural networks, and we find that global strategies such as Shapley values can be effectively used in practice. Our analysis also highlights that Shapley values may help choose the network architecture, as we find that fully-connected neural networks perform better than convolutional neural networks  in predicting and interpreting the Heston model prices to parameters relationship. 

\end{abstract}

\bigskip

\noindent {\bf AMS classification codes:} 68T07, 91G20, 91G60.

\newpage
{\small \tableofcontents}
\vfill
{\footnotesize \noindent The opinions here expressed  are solely those of the authors and do not represent in any way those of their employers.}
\newpage

\pagestyle{myheadings} \markboth{}{{\footnotesize  Brigo, Huang, Pallavicini, Oc\'ariz, Interpretability in DL for Volatility Smile Calibration}}

\section{Introduction}
\label{sec:introduction}

In recent years, machine learning has experienced a surge in popularity and its applicability has been extended to almost every industry. In particular, the financial sector is no stranger to machine learning. Tools such as deep learning are already used in several applications including portfolio management, algorithmic trading, cryptocurrency and blockchain studies, fraud detection, and model calibration as described by~\cite{Ozbayoglu2020DeepLF}. In this project we will focus on deep learning for investigating the relationship between a volatility model and the related volatility smile. In other terms, given the volatility smile produced by a model, we wish to find the model parameters that generated that smile. When applied to volatility smiles obtained from market data, this may be helpful in  model calibration. With slight abuse of language we refer to our inversion problem as ``calibration'' even if we do not work with real market data calibrations in this paper. We focus on the \cite{Heston1993ACS} model, which is one of the reference models in derivative option pricing. This model is well understood both in theory and in practice. Several contributions regarding using Artificial Neural Networks (ANNs or NNs for short) for model calibration already exist in the literature. This topic has been explored by several authors such as~\cite{mcwnn, bayer2018deep, Horvath2019DeepLV, Bloch2019OptionPW, roeder2020volatility} and more. We also refer the reader to \cite{ruf2020neural} for a review on machine learning for option pricing and hedging, where a section is devoted to the calibration problem. Yet, none of these contributions explores interpretability of the results.

Interpretability methods are conceived as a response to the black box nature of NNs, where the functioning of the networks and the input-output relationship are difficult to understand, due to non-locality and non-linearity. See for example~\cite{brigo2019DLI} for an overview. In our context, although the NN aims at learning the model calibration, the NN approximation will not give the user any insight on the form of the original function. There is no straightforward link between the weights and the function being approximated. This can be a relevant concern, since the model may be learning the wrong model features but still giving satisfactory results within the available data set. This is easier to illustrate with an image recognition example. Imagine we build a network that aims at identifying cats, but all the cats in our data set are wearing a collar. The network may learn that having a collar is what characterises cats, which is obviously not true. Using interpretability we could understand which input features affect the output the most and identify the problem. This is of paramount importance in quantitative finance since decisions that involve large amounts of funds are being made based on complex machine learning algorithms, which can be extremely hard to understand and in which errors can go unnoticed. As claimed by~\cite{molnar2019}, there is no specific mathematical definition of interpretability. One (non-mathematical) definition was proposed by~\cite{Miller2019ExplanationIA}: ``Interpretability is the degree to which a human can understand the cause of a decision.'' Another one was by~\cite{kim2016examples}: ``Interpretability is the degree to which a human can consistently predict the model's result''. 


These initial definitions may be helpful but they do not fully address what we intuitively understand to be interpretability. Reaching a precise and encompassing definition is certainly work in progress. Interpretability methods are a hot topic in the machine learning literature, but to the best of our knowledge, we have found only a few examples on the subject in the financial literature. In particular we refer to~\cite{wang2019alphastock}, where a sensitivity analysis is applied to a  trained NN to model a trading strategy, to~\cite{Demajo20} for applications on credit scoring, and to~\cite{Moehle21} for applications in performance attribution for portfolio analysis. Moreover, in~\cite{forecasting_damiano} interpretability is briefly considered for recovery-rate predictions in non-performing loans, but without using deep NNs.

In this paper we will analyse the applicability of global and local interpretability methods to our calibration problem and we will calibrate the Heston model with two different NN architectures. We will compare the global and local interpretability results, test their applicability, reliability and consistency, and investigate results on both networks. Since the Heston model is well understood, theoretical and empirical facts about the model will serve as a solid benchmark for our results, so that we can discriminate the best NN architecture to be used in the analysis. On the other hand, this will show which methods work best and could potentially be applied to more complex and less understood models.
Our main finding from the Heston case is that interpretability strategies based on global tools from cooperative game theory, such as Shapley values, may outperform interpretability coming from local tools where one fits simple models locally to the network input-output relationship. Given that this relationship is non-linear and non-local, applying local methods, often based on linearisation, is clearly not ideal in interpreting the NN behaviour. This is confirmed by our findings. We also find that local interpretability does not align with our financial intuition  on the Heston model, while global interpretability and Shapley values in particular do. Moreover, we can use Shapley values as a practical tool to discriminate among NN architectures and to select the ones that better match the model behaviour by establishing clear correspondences between calibrated options and model parameters. 
Indeed, we find that a Fully-Connected NN performs better than a Convolutional NN, contrary to what happens with image recognition. Secondly, we can use these tools to investigate the model's behaviour when we are confident of the NN architecture but we lack a clear model interpretation.

\medskip

The structure of the paper is as follows. In Section~\ref{sec:interpr} we present the problem of using interpretability methods in calibrating pricing models via a NN, along with a description of the main tools at our disposal. In Section~\ref{sec:heston} we proceed by illustrating how to calibrate the Heston model by means of a NN. In particular, we describe two NN architectures: a Fully-Connected NN and a Convolutional NN. Then, in Section~\ref{sec:discussion} we apply the interpretability tools to our case, and we present and discuss our results. In Section~\ref{sec:conclusion} we review our contributions and give hints for further developments.

\section{Interpretability of neural network calibration}
\label{sec:interpr}

In the following section we present the relevance of NNs in the calibration of pricing models and select the specific approach we are going to use for the interpretability analysis, whose tools and methods are reviewed in the subsequent sections.

\subsection{Pricing-model calibrations via neural networks}
\label{sec:pricing-model}

When a pricing model is conceived, the performance of the calibration process is of paramount importance. One is given liquid market prices of some benchmark products, typically options, and has to find the parameters of the model that generate model prices as close as possible to the given market prices according to a chosen criterion. This involves minimizing a calibration error, measuring the discrepancy between the market prices and the model prices.  Calibrating model parameters to liquid market quotes can be very time
consuming, especially if there is no analytic solution for pricing the benchmark products. If the calibration is a global one, that is, if we seek to find a global optimum, the process becomes even slower. A possible way to deal with this issue, and ensuring a fair level of accuracy, speed, and robustness, is using a NN as part of the calibration to speed up the process.

We can better understand the mechanism if we consider the procedure proposed in \cite{Horvath2019DeepLV}. The authors propose a two-step algorithm for the calibration of the so called rBergomi model, a rough volatility model. Interestingly, the proposed approach is generic and not limited to this specific model. In the first step the authors build a NN to learn the pricing map from model parameters to market quotes, then the calibration is performed by using the trained NN to efficiently find the market quotes from the model parameters. This map is now very fast, and it is easier to invert in a second step with standard optimization techniques. The authors study both gradient-based and gradient-free methods. We may refer to this method as a two-step procedure. We stress that in this approach the NN is used only to obtain a faster version of the pricing map, without learning the whole calibration procedure.



On the other hand, we could take a different approach and use the NN to learn the whole calibration procedure. In \cite{roeder2020volatility} the authors focus on both the Heston and the rBergomi model, and they build a NN to learn the map from market quotes to model parameters, so that the whole calibration process is learnt by the NN. The same approach is also described in \cite{mcwnn} for the Hull and White model. In our analysis we choose to follow this procedure since the resulting network has to learn the whole calibration process, leading to a more challenging scenario for our interpretability methods.




\subsection{Interpretability models and theoretical background}
\label{sec:localvsglobal}

Interpretability aims at giving the user some insight into the machine learning algorithm and how this is making decisions and learning the input and output relationships. In our pricing-model calibration problem, interpretability allows us to understand which inputs affect the most our model parameters.

These tools can be useful in two different situations. First, if we have complete knowledge of our model, as in the Heston case, we can test if such correspondence between input data and model parameters matches our intuitive understanding of the pricing model. If this is not the case then we could reject the NN architecture in favour of more suitable choices. Second, if we lack knowledge of the model, we can use these tools precisely to improve our understanding of the model's behaviour.

In \cite{chakraborty2017interpretability} a survey of prior work on interpretability in deep learning models is presented alongside an introduction of relevant concepts mainly on two topics: model transparency and model functionality. For further details, \cite{molnar2019} published a book integrating theoretical background and implementation examples of current interpretable methods. This book is helpful for readers to build an overall framework of how to make deep learning models interpretable. However, it may happen that some NN models can produce accurate predictions with poor interpretability. In \cite{sarkar2016accuracy} the trade-offs between accuracy and interpretability in machine learning are discussed. In that paper the authors propose the TREPAN algorithm to draw better performing decision trees from a NN that balances prediction accuracy and tree interpretability. We would like to highlight the difference between algorithm transparency and interpretability methods. Algorithm transparency looks at the algorithm itself, that is, how it works and what kind of  relationships it applies to, not the specific model or prediction making process. On the other hand, that paper focuses on interpretability, which requires knowledge of the algorithm and the data, and analyses the trained model. Interpretability is less understood than algorithm transparency and it is an area of active research.

Many proposals for interpretability methods are published in the literature. We cite a few to give an overview of recent developments in the field\footnote{Note that LIME and SHAP, can be implemented within the \texttt{DeepExplain} framework, whose Python package can be found at \url{https://github.com/marcoancona/DeepExplain}. The individual Python packages for LIME and SHAP are available at \url{https://github.com/marcotcr/lime} and \url{https://github.com/slundberg/shap}, respectively.}. In \cite{bach2015pixel} the authors develop LRP (Layer-Wise Relevance Propagation) to interpret classifier predictions of automated image classification, and they provide a visualization tool that can draw the contributions of pixels to the NN predictions. LIME (Local Interpretable Model-agnostic Explanations) is proposed by \cite{lime}, which is able to explain the predictions of any classifier or regressor. Specifically, it is a local surrogate model since it trains a local explanation model around individual predictions. In \cite{Shrikumar2016NotJA} and \cite{Shrikumar2017LearningIF} the authors introduce DeepLIFT (Learning Important FeaTures) and {Gradient*Input} to attribute feature importance scores by evaluating the difference between the activation of every neuron and the corresponding reference level and they conclude that DeepLIFT significantly outperforms gradient-based methods. In \cite{Lundberg2017AUA} the authors modify classical Shapley values into a unified approach that can interpret model predictions globally and they call it SHAP (SHapley Additive exPlanations) values. In \cite{Sundararajan2017AxiomaticAF} the authors present Integrated Gradients, a gradient-based attribution method driven by sensitivity and implementation invariance. In \cite{ancona2018towards} the authors experiment with multiple interpretability methods such as gradient-based attribution methods (Saliency maps, {Gradient*Input}, Integrated Gradients, DeepLIFT, and LRP) and perturbation-based attribution methods (Occlusion and Shapley Value sampling), where Occlusion is based on the work by~\cite{Zeiler2014VisualizingAU}.

We can classify interpretability models in two main categories according to~\cite{molnar2019}: global and local interpretability methods. Global interpretability methods aim at recognizing how the model makes decisions based on a holistic view of its features and each of the learned components such as weights, hyperparameters, and the overall model structure. Global model interpretability helps to understand the distribution of the target outcome based on the features. In contrast, local interpretability focuses on a single model prediction at a time. If we analyse an individual prediction, the behaviour of the otherwise complex model might appear more pleasant, although this does not have to necessarily be the case. Locally, the prediction might only depend linearly or monotonically on some features, rather than having a complex dependence on them. Also, being deep NNs non-linear and non-local, it seems unlikely that local methods can give any relevant insight on the network behaviour at large. In the previous list most methods are local except for the methods based on Shapley Values.




\subsection{Local interpretability}
\label{sec:localim}

When a deep learning model is used to make predictions, it is often asked what is the logic behind the model and how the features contribute to the predictions. Local surrogate models, to some extent, answer this question. They interpret individual predictions of a NN based on some feature. Mathematically, let $f$ be the original model that we want to explain, and let $\hat{f}$ be the prediction estimated by a deep NN. We aim to explain a single prediction $y=\hat{f}(x)$ based on the input $x$. In interpretability models, instead of using the original input $x$, the simplified input $x'$ is often used, which is defined by a mapping function  $x=h_x(x')$. With original input $ x $ fixed, the simplified input $ x'$ sets each component to binary, where 1 means that the input component is present, and 0 means absent. The ``absence'' here means that the feature value is equal to the mean of the data set. 

Given the input $x$, this has some components $j$ that are absent, such that $x'_j=0$, and other components $k$ that are present, such that $x'_k=1$. For example, we have
$$
x=[m_1,v_1,m_2,v_2,v_3] \Longrightarrow  x'=[0,1,0,1,1],
$$
where $m_i$ are the averages of the input data $x$ and $v_i$ are other different values.
Then, $h_x$ enables us to map $x'$ back to the original input $x$,
$$
h_{[m_1,v_1,m_2,v_2,v_3]}([0,1,0,1,1]) = [m_1,v_1,m_2,v_2,v_3].
$$
Next, an explanation model $g$ is called \textit{local} if $\hat{f}$ can be ideally approximated by $g$, that is, for $z' \approx x'$
$$ 
g(z')\approx \hat{f}(h_x(z')).
$$
For such a explanation model $g$ to have \textit{additive feature attribution}, $g$ can be written as an additive structure
\begin{equation}\label{eq:AFA}
    g(z')=\phi_0 + \sum_{i=1}^{M}\phi_i z'_i,
\end{equation}
where $M$ is the number of simplified input features, $z'\in \{0,1\}^M$ is the coalition vector, and $\phi_i \in \mathbb R$ is the $i$-th feature attribution. Note that $\phi_0$ is the case when $z_i'=0$ for $i=1,\ldots,M$. In other words, each $z'_i$ is equal to its respective data set mean. $g(z')$ accounts for the contribution of the different feature attributions $\phi_i$ and aims at approximating the model output $f(x)$.

\cite{molnar2019} summarised the process in the following steps:

\begin{itemize}

    \item Select an instance $x$ whose model prediction is to be explained.
    
    \item Perturb your data set and generate model predictions for these new data points.
    
    \item Set weights for the new samples based on their proximity to the instance $x$.
    
    \item Train the explanation model $g$ with weights on the data set with variations. 
    
    \item Interpret the obtained prediction by explaining the local model $g$.
    
\end{itemize}

LIME, DeepLIFT and LRP are local surrogate models that use equation~\eqref{eq:AFA} locally in $x$ to fit $\phi$ and get the explanations. We give a more detailed explanation of these methods in Appendix~\ref{sec:lmdetail}.

\subsection{Global interpretability}
\label{sec:globalim}

According to~\cite{lipton2017mythos} we can describe a model as interpretable if we are able to comprehend the entire model at once. This appears to be a daunting task for a complex NN that is both non-linear and non-local. For this purpose, we can use global interpretability methods. We focus on interpretability methods based on Shapley values.

We start by introducing the foundation behind classical Shapley values. Shapley values originated from cooperative game theory and were coined by~\cite{shapley1953value}. The idea is to compute the average marginal contribution of each player $i$ to the total game gain (payout) across all possible collaborating players. 

Shapley values can be used for interpretability in deep learning to explain the predictive model. In this context, we simply view the prediction task as a game, and each feature value as a player in the game. Subsequently, the gain of this task is the difference between the actual prediction for a particular instance and the average prediction for all instances. Note that the Shapley value is not equivalent to the difference in the prediction when the feature is removed from the model. It focuses on how to fairly distribute total gains among the features. As an explanation model, Shapley values has the additive feature attribution structure as~\eqref{eq:AFA},

\begin{equation*}
g(z')=\phi_0 + \sum_{i=1}^{M}\phi_i z'_i.
\end{equation*}%
Specifically in the case of Shapley values,
\begin{equation*}
      \phi_{i}(\hat{f},x)=\sum_{\substack{z' \subseteq x'\backslash \{i\}}}\frac{|z'|!(M-|z'|-1)!}{M!}\left( \hat{f}\left( h_x(z' \cup \{i\}\right))-\hat{f}\left(h_x(z')\right)\right).
\end{equation*}%

Intuitively, feature values $(x_k)_k$ enter in random order, and all values contribute to the prediction. The Shapley value of $x_i$ is the average (on orders) of change in the prediction when $x_i$ joins. In deep NNs the four Shapley value properties become the following:

\begin{itemize}

\item[(i)]  Efficiency: total gain is recovered, which means the sum of feature contributions must be equal to the prediction for an instance $x$ minus the average of all possible instances. Mathematically,
\begin{equation*}
    \sum_i \phi_i(\hat{f},x) = \hat{f}(x)-E[\hat{f}(x)].
\end{equation*}%

\item[(ii)]  Dummy: if a feature $j$ never adds marginal value, its Shapley value $\phi_j = 0$. For a coalition of features $S$, the prediction remains the same when $x_j$ joins 

\begin{equation*}
    f(S\cup \{x_j\}) = f(S) \quad \text{ for all } S\subseteq \{x_1,\ldots,x_M\},
\end{equation*}%
then
\begin{equation*}
    \phi_j = 0.    
\end{equation*}%

\item[(iii)]  Symmetry: any two features contributing equally to the total gain have the same $\phi$. In other words, for all $S\subseteq \{x_1,\ldots,x_M\} \backslash \{x_k,x_j\}$, if 
\begin{equation*}
    f(S\cup \{x_k\}) = f(S\cup \{x_j\})
\end{equation*}
then
\begin{equation*}
    \phi_k = \phi_j.
\end{equation*}%

\item[(iv)]  Additivity: if a prediction task is composed of $f_1$ and $f_2$, then the Shapley values are $\phi(f_1) + \phi(f_2)$.

\end{itemize}

SHAP (SHapley Additive exPlanations) was proposed by~\cite{Lundberg2017AUA} and it is based on optimal Shapley Values. It also aims to calculate each feature's importance to the prediction of an instance $x$. The authors built SHAP as a unified measure of feature contribution, and some methods to effectively generate them such as: Kernel SHAP (Linear LIME + Shapley values), Tree SHAP (Tree explainer + Shapley values), and Deep SHAP (DeepLIFT + Shapley values). Additionally, SHAP combines many global interpretation methods that are based on aggregations of Shapley values. 

SHAP admits the additive feature attribution structure as equation~\eqref{eq:AFA}, a linear model of binary variables
\begin{equation*}
g(z')=\phi_0 + \sum_{i=1}^{M}\phi_i z'_i.
\end{equation*}%
Moreover, SHAP considers the case when all the features are present. For the simplified input $x'$ of an instance $x$, all components of the vector $x'$ are present, being 1. Then equation~\eqref{eq:AFA} becomes:

\begin{equation}
g(x')=\phi_0 + \sum_{i=1}^{M}\phi_i .
\end{equation}%

Apart from Efficiency, Dummy, Symmetry and Additivity,~\cite{Lundberg2017AUA} pointed out another three desired properties that SHAP should satisfy. The three additional properties read as follows:

\begin{itemize}

\item[(1)] Local accuracy: this property requires that the original model $f$ with an instance $x$
can be at least estimated by the explanation model $g$ with the corresponding simplified input $x'$.
\begin{equation*}
    f(x) = g(x')=\phi_0 + \sum_{i=1}^{M}\phi_i x'_i
\end{equation*}%
where $x=h_x(x')$.

\item[(2)] Missingness: since the binary components in the simplified input $x'$ are either present or absent, features missing in the original input $x$ should have no impact. If $x'_i=0$, then $\phi_i = 0$.

\item[(3)] Consistency: when a model is changed, the change of the input's attribution should be consistent with the change of its simplified input's contribution. Denote $f_x(z')=f(h_x(z'))$, and define $z'\backslash \{i\}$ by setting $z_i'=0$. For any two models $f_1$ and $f_2$ satisfying
\begin{equation*}
    f_{2x}(z') - f_{2x}(z'\backslash \{i\}) \geq f_{1x}(z')-f_{1x}(z' \backslash \{i\}),
\end{equation*}%
then for all inputs $z'\in \{0,1\}^M$, their effects are consistent with such a change
\begin{equation*}
    \phi_i(f_2,x) \ge \phi_i(f_1,x).
\end{equation*}%

\end{itemize}

The authors also claimed that there exists only one possible explanation model $g$ satisfying the additive feature attribution structure and the SHAP properties (1) to (3) depicted above:

\begin{equation}
    \phi_{i}(f, x)=\sum_{z^{\prime} \subseteq x^{\prime}} \frac{\left|z^{\prime}\right| !\left(M-\left|z^{\prime}\right|-1\right) !}{M !}\left[f_{x}\left(z^{\prime}\right)-f_{x}\left(z^{\prime} \backslash \{i\}\right)\right]
\end{equation}%
where the number of non-zero components in vector $z'$ is denoted by $|z'|$. 

In the case of SHAP values, $f_x(z')=f(h_x(z'))=\mathbb{E}[f(z)|z_A]$ and $h_x(z')=z_A$, where $A$ is the set of non-zero indices in $z'$ and $z_A$ has missing values for the features in the set $A$. Additionally, we use $\bar{A}$ to refer to the set of features not present in $A$. For example, start from the base value $\mathbb{E}[f(z)]$ (corresponding to $\phi_0$),  SHAP attributes $\phi_1$ to $\mathbb{E}[f(z)|z_1=x_1]$, the change in the expected model prediction conditional on the feature $1$, $\phi_2$ to $\mathbb{E}[f(z)|z_{1,2}=x_{1,2}]$, $\phi_3$ to $\mathbb{E}[f(z)|z_{1,2,3}=x_{1,2,3}]$, and so on. Finally, the original model $f(z_A)$ can be approximated by a conditional expectation $\mathbb{E}[f(z)|z_A]$. 
Therefore, SHAP values are actually those Shapley values of a expectation of the original model $f$ conditional on $z_A$.

Although it is challenging to compute exact SHAP values, they can still be approximated by combining current additive feature attribution methods according to \cite{Lundberg2017AUA}. Two optional assumptions, model linearity and feature independence, are used to simplify the expectation calculation:
\begin{equation} \label{eq:SHAP}
\begin{split}
f\left(h_{x}\left(z^{\prime}\right)\right) &=E\left[f(z) \mid z_{A}\right] \\ &=E_{z_{\bar{A}} \mid z_{A}}[f(z)] \\ & \approx E_{z_{\bar{A}}}[f(z)] \\ & \approx f\left(\left[z_{A}, E\left[z_{\bar{A}}\right]\right]\right) .
\end{split}
\end{equation}%

The first equation in \eqref{eq:SHAP} is obtained by simplifying the input mapping. Next, the conditional expectation taken over $z_{\bar{A}} | z_{A}$ is calculated, instead of the marginal expectation and the value is approximated assuming feature independence. Lastly, to obtain the final result, the model is assumed to be linear. For more information refer to~\cite{Lundberg2017AUA} and~\cite{molnar2019}.

\section{A detailed example with the Heston model}
\label{sec:heston}

We run the subsequent analysis on the Heston model presented in \cite{Heston1993ACS}, because it is often used in practice and it is well understood. This provides us with a benchmark to evaluate which interpretability methods give satisfactory results and could be used in less understood models in the future.

\subsection{A brief overview of the Heston model}
\label{sec:heston_overview}

The original Heston model is presented under the physical measure $\mathbb{P}$, also known as the objective measure. Yet, we move on to the risk-neutral measure $\mathbb{Q}$ to present our analysis as this is more convenient from the point of view of option pricing and calibration.

We define $S_t$ as the asset price and $v_t$ as its instantaneous variance. $v_t$ can be viewed as a mean reverting square-root process. The system of stochastic equations of the Heston model under the risk-neutral measure $\mathbb{Q}$ is written as 
\begin{align*}
&d S_t = \sqrt{v_t} S_t dW_{1t},\quad S_0\\
&d v_t = \kappa (\theta - v_t) dt + \sigma \sqrt{v_t} dW_{2t},\quad v_0\\
&dW_{1t} dW_{2t} = \rho dt
\end{align*}
where $W_{1t}$ and $W_{2t}$ are the Brownian motions with correlation $\rho$. We set the risk-free rate and the dividend yield equal to zero for simplicity, although we could potentially add them in a straightforward way.

All the free parameters are collected in the set $\psi := \{v_0, \rho,\sigma, \theta, \kappa\}$. We list below the range and meaning of the parameters.
\begin{itemize}
  \item $v_0>0$ is the initial value of the variance.
  \item $\rho\in [-1,1]$ is the correlation between the two Brownian motions, or the instantaneous correlation between asset and instantaneous variance.
  \item $\sigma\ge 0$ is the volatility of the volatility.
  \item $\theta>0$ is the long-term mean of the instantaneous variance of the asset price. 
  \item $\kappa\ge 0$ is the speed at which $v_t$ reverts to $\theta$. 
\end{itemize}

It can be seen that this model precludes negative values for $v_t$, and when $v_t$ tends to zero it becomes positive. The mean reversion property of $v$ is important. Moreover, if the parameters satisfy the Feller condition
\begin{equation}\label{eq:feller}
    2\kappa\theta > \sigma^2,
\end{equation}%
then the instantaneous variance $v_t$ is strictly positive. In the following section, we impose the Feller condition when generating our synthetic data. Notice that when the calibration is done in industry, the Feller condition is sometimes relaxed to obtain a better fit of the market data. The Heston model is well studied in the financial literature, and we refer to \cite{Gatheral06} for details.

\subsection{Calibration via neural networks}
\label{sec:Calibration via neural networks}

When calibrating the Heston model we aim at finding the optimal model parameters $\psi^\star$ such that the model prices best match the market data. This is effectively a non-linear constrained optimization problem, where we minimise the error between market option quotes ($C^{\rm mkt}(T,K)$) and their prices calculated by the Heston model ($C(\psi;K,T)$) for a selection of maturities $T$ and strikes $K$ quoted by the market in the set $\{(T_i,K_i) : i=1\ldots n\}$.
\begin{equation}
  C(\psi;K,T) = {\mathbb E}^{\mathbb{Q}(\psi)}[(S_T-K)^+]
\end{equation}%
The calibration of the parameters to the observed market data consists in minimizing the loss $L$ with respect to the model parameter set $\psi$.
\begin{equation}
  L(\psi;C^{\rm mkt}) := \sum_{i=1}^n \left(C(P;K_i,T_i) - C^{\rm mkt}_i \right)^2
\end{equation}%

The solution of the equation above can be regarded as a function of the market data mapping to the domain where the model parameters reside.
\begin{equation}
\label{eq:calibmap}
  C^{\rm mkt} \longrightarrow \psi^\star := \argmin_\psi L(\psi;C^{\rm mkt})
\end{equation}%

As we discussed in Section~\ref{sec:pricing-model} we proceed as in \cite{roeder2020volatility}, by using a NN to approximate the calibrated model parameters function given by Equation~\eqref{eq:calibmap}.



\subsection{Description of the synthetic data set}
\label{sec:heton_calib}


We build our data set in a synthetic way. We do not start from real market quotes since we are only interested in interpreting the NN used to learn the calibration procedure. We suggest training the NN with real market data in future work.

Regarding the following steps, we refer to the work by \cite{bayer2018deep} for the calibration setup. The features for the NN we are going to construct to approximate the calibration map given by Equation~\eqref{eq:calibmap} are the volatilities of market plain-vanilla options $C^{\rm mkt}$, while the labels are the model parameters, $\psi$.

First, we start by randomly generating the Heston model parameters $\psi$ for labels with large enough size, and we calculate the corresponding implied-volatility surface for the input features. We randomly generate each parameter using a uniform distribution within the bounds defined in Table~\ref{tab:para_bounds} according to previous work by~\cite{Horvath2019DeepLV} and~\cite{roeder2020volatility}. Once the model parameters are generated we select the sets satisfying the Feller condition given by Equation~\eqref{eq:feller}. We disregard possible correlations between model parameters when we generate them, since we prefer to assume a null prior knowledge of the joint distribution as suggested in~\cite{bayer2018deep}. For the main discussion we use a data set of size $10^{4}$ in line with the order of magnitude of other data sets used by~\cite{Horvath2019DeepLV} and~\cite{roeder2020volatility}. Additional results using a larger data set of size $10^{5}$ can be found in Appendix \ref{largerdataset}, although little improvement is found at the expense of a high computational cost.

\begin{table}
\centering
\begin{tabular}{|l|c|c|c|c|c|}
	\hline
	Parameter & $v_0$ & $\rho$ & $\sigma$ & $\theta$ & $\kappa$ \\
	\hline
	Lower bound & 0.0001 & -0.95  & 0.01 & 0.01  & 1.0 \\
	\hline
	Upper bound & 0.04 & -0.1  &  1.0 &  0.2  & 10.0\\
	\hline
\end{tabular}
\caption{Parameter bounds for the Heston model}
\label{tab:para_bounds}
\end{table}

Next, we evaluate by means of the Heston model the plain-vanilla option prices on a grid of maturities and strikes given by
\begin{equation*}
  K = \{0.5,0.6,0.7,0.8,0.9,1.0,1.1,1.2,1.3,1.4,1.5\}
\end{equation*}%
\begin{equation*}
  T = \{0.1,0.3,0.6,0.9,1.2,1.5,1.8,2.0\}
\end{equation*}
where we assume that the spot price $S_0$ is equal to 1. Thus, the dimensions of the resulting volatility matrices are $8$ strikes per $11$ maturities, leading to a feature collection of $88 = 8\times 11$ volatilities, while the corresponding labels are the $5$ model parameters.

\begin{figure}
    \centering
    \includegraphics[width=.45\textwidth,keepaspectratio]{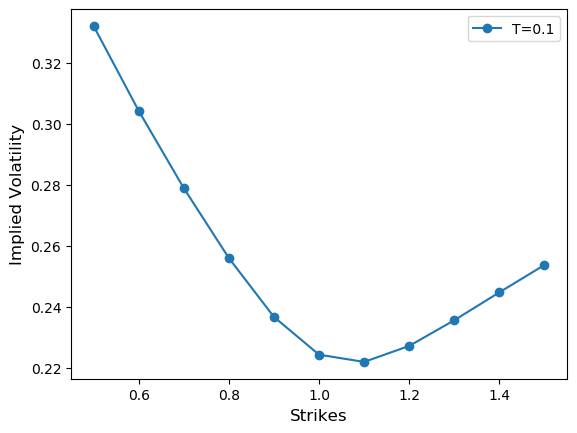}
    \hspace*{1em}
    \includegraphics[width=.45\textwidth,keepaspectratio]{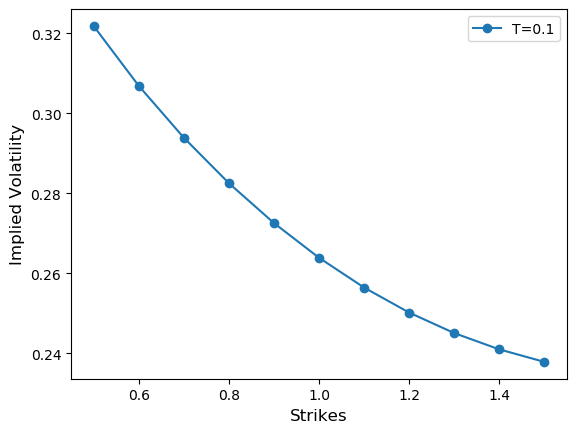}
    \caption{Different shapes for the volatility smile (option maturity is 0.1 years). Left panel: an instance from the test input data set. Right panel: an instance from the train input data set.}
    \label{fig:vol_smile}
\end{figure}

The data set of size $10{,}000$ is split into a training set of $8{,}500$ data points and a testing set of $1{,}500$ data points. Taking a closer look at the data if we, for example, select one volatility matrix of the input data in the testing set and fix the maturity to be $T=0.1$, the volatility smile is shown as an asymmetric U-shape in the left panel of Figure~\ref{fig:vol_smile}. However, in the case of the volatility smile at another entry of the training set with fixed maturity $T=0.1$, as shown in the right panel of the same figure, the smile is monotonic. Therefore, our synthetic data does not guarantee a specific shape of the volatility smile.

It is well-known that substantial differences in the scale of the input data can lead to difficulties while training and cause instabilities. Therefore, it is common practice to standardize the input data to numerically aid the training of the NN. A Fully-Connected Neural Network (FCNN) and a Convolutional Neural Network (CNN) are implemented in the papers by \cite{Horvath2019DeepLV} and \cite{roeder2020volatility}, respectively. For the FCNN the data is scaled in a range from 0 to 1 before being inputted to the network, while for the CNN the data is scaled to have a mean of 0 and a variance of 1. In addition, after scaling the input data of the FCNN, a whitening process is applied. Whitening is only applied to the FCNN case since empirically it does not seem to aid training in the case of the CNN. In particular, we rely on ZCA-Mahalanobis whitening, described by~\cite{whitening}, whose aim is to de-correlate the input matrices by multiplying the centred input data by a de-correlation matrix. In this way, the data is linearly transformed so that the sample correlation matrix of the training data is the identity. This whitening process presents the advantage that the de-correlation is achieved by only applying minor changes to the original input.

\subsection{Neural network architectures}
\label{sec:heston_nn}

The NNs here presented are inspired by~\cite{Horvath2019DeepLV} and~\cite{roeder2020volatility}. Two NN architectures are considered: the FCNN and the CNN. Both architectures are Feedforward Neural Networks because there are no feedback connections in which the outputs of the model are fed back into themselves. Information flows from the input to the intermediate computations that define the function and, then, to the output. FCNNs are called Fully-Connected because all their nodes in each layer are connected to all the nodes in the previous and subsequent layers. While they are considered to be a rather simple NN architecture, they give good and fast calibration results. CNNs derive their name from their characteristic convolutional layers. The primary purpose of the convolution is to highlight features from the input. CNNs were inspired by biological processes in that the connectivity pattern between neurons resembles the human visual cortex. Individual neurons respond to stimuli only in a specific region of the visual field, which is known as the receptive field. CNNs are commonly applied to analyzing visual imagery.

We start by calibrating the model with the FCNN and we show that this simple model is enough to obtain good results. Next, we test the CNN and we compare the results with the original simpler architecture. The CNN is implemented to see whether it improves the calibration, since it has previously outperformed the FCNN in a number of applications and has also been implemented by researchers for volatility smile calibration. Also, since FCNNs and CNNs learn the calibration in different ways, it is interesting to apply the interpretability methods to both networks.

\begin{figure}
    \centering
    \includegraphics[scale=0.4]{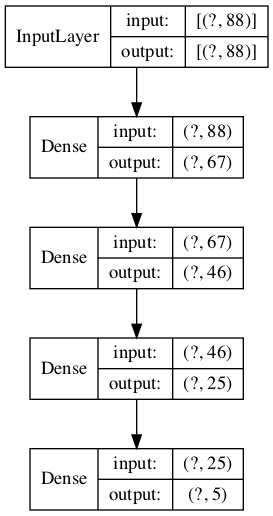}
    \caption{FCNN architecture summary. All layers have two dimensions. The first being the batch size, which is represented by the symbol: ``{?}'', left unspecified within the network architecture and it is defined during training, and the second being the data size. Remember that in the case of the FCNN we flatten the input $8\times11$ volatility matrix into an array with 88 entries. We decrease the number of nodes until the output only has 5, one for each of the model parameters.}
    \label{fig:F_model_plot}
\end{figure}

The FCNN here presented consists of four fully-connected dense layers as shown in Figure \ref{fig:F_model_plot}. The hidden layers use ELU as their activation function and the output layer uses a hard sigmoid. The output layer generates a five output vector corresponding to the Heston model parameters. Four layers with a decreasing number of neurons are found to be enough to obtain excellent results. $10{,}396$ trainable parameters are required. In the case of the FCNN the mean-squared logarithmic error is used for the loss function, since it seems to be the best suited for this architecture after contrasting its performance with other typical losses. The Adam optimiser is used for training.

\begin{figure}
    \centering
    \includegraphics[width=1.0\textwidth,keepaspectratio]{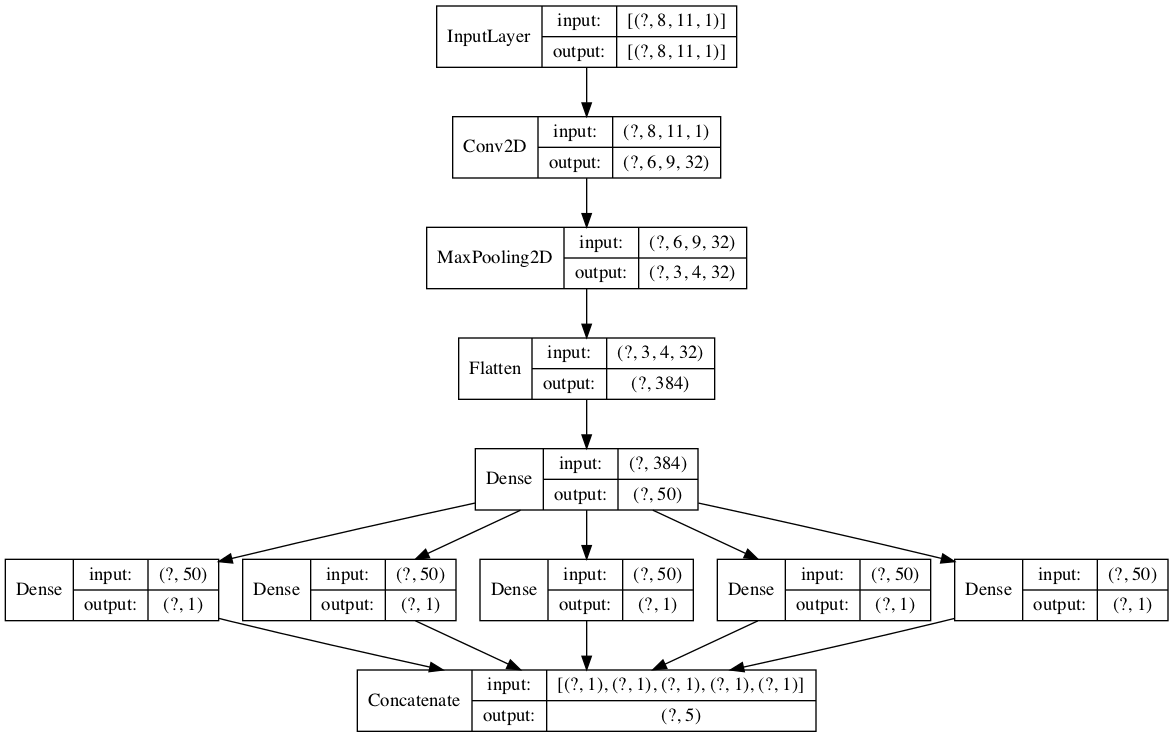}
    \caption{CNN architecture summary. The first dimension in all the layers: ''$?$'' refers to the batch size. It is left as an unknown or unspecified variable within the network architecture so that it can be chosen during training. All layers before flattening have four dimensions. The second and third dimension correspond to the matrix dimensions. As we can observe, before the convolution, the input matrix has dimensions $8\times11$ as previously explained. The last dimension corresponds to the channel dimension. In the $InputLayer$ this takes a value of 1, since we are just inputting a two dimensional matrix. Conv2D uses 32 filters and generates 32 activations from the original one dimensional input. Hence, the last dimension expands to 32 after Conv2D. After flattening the data, we are left with two dimensions, the first being the batch size as before, and the second being the length of the array. Lastly, we code a common dense layer and 5 different layers to predict each of the model parameters. The last layer simply concatenates the prediction of each of the model parameters into one array.}
    \label{fig:C_model_plot}
\end{figure}

In the case of the CNN the ELU activation function is used for the hidden layers too. The CNN starts with an input layer which has no parameters to be calibrated but it is rather created to define the input shape of the training data. It is referred to as ``InputLayer'' in Figure~\ref{fig:C_model_plot}. After that, a two dimensional convolutional layer with $3 \times 3$ filters is used. The convolutional layer is meant to extract features from the input and aims at highlighting the most important attributes of the data. After testing different filter sizes the $3 \times 3$ filters proved to perform best. The convolutional layer uses 32 filters. Next, maxpooling is applied, which again aims at highlighting the most important features with a $2\times 2$ kernel. Maxpooling halves the dimension of the convolutional layer output. After that, the output is flattened into a $3\times4\times32 = 384$ tensor and the fully-connected part of the NN takes on. The flattened data is passed onto a dense layer which outputs a vector of just 50 entries. Lastly, we concatenate 5 different layers with a single output, one for each of the five Heston model parameters that need to be calibrated and make use of custom activation functions that improve the performance of the NN. The custom activation functions are hyperbolic tangent functions which have been normalised for each of the  model parameters. These activation functions were chosen based on the research by~\cite{Horvath2019DeepLV}. This architecture equates to a total of $19{,}825$ trainable parameters. When training the model the root mean square error is used as the loss and the Adam optimisation algorithm is implemented.

\subsection{Calibration results}
\label{sec:calib_result}

\begin{figure}
    \centering
    \includegraphics[width=1.0\textwidth,keepaspectratio]{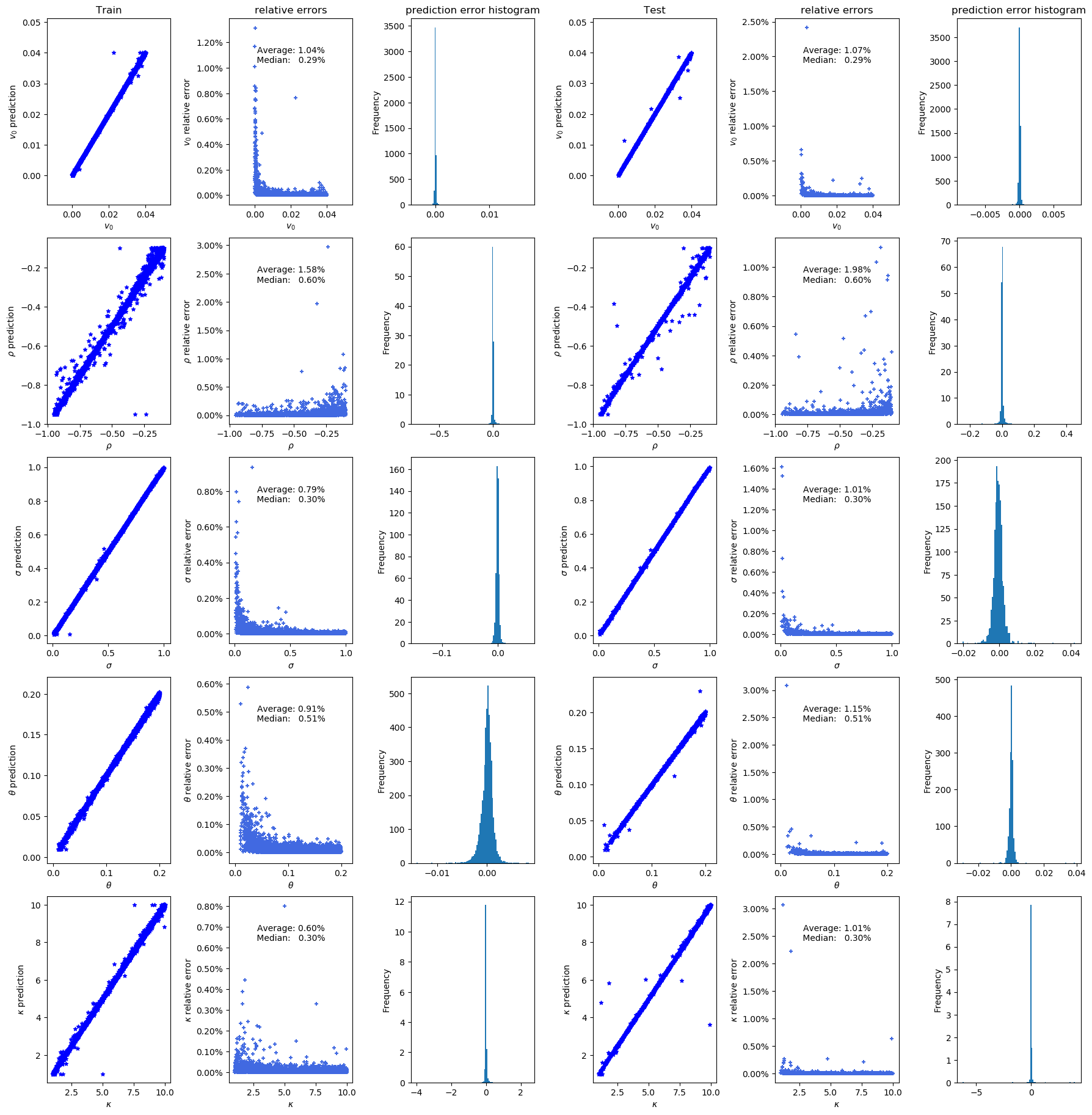}
    \caption{FCNN prediction errors.}
    \label{fig:F_err}
\end{figure}

\begin{figure}
    \centering
    \includegraphics[width=1.0\textwidth,keepaspectratio]{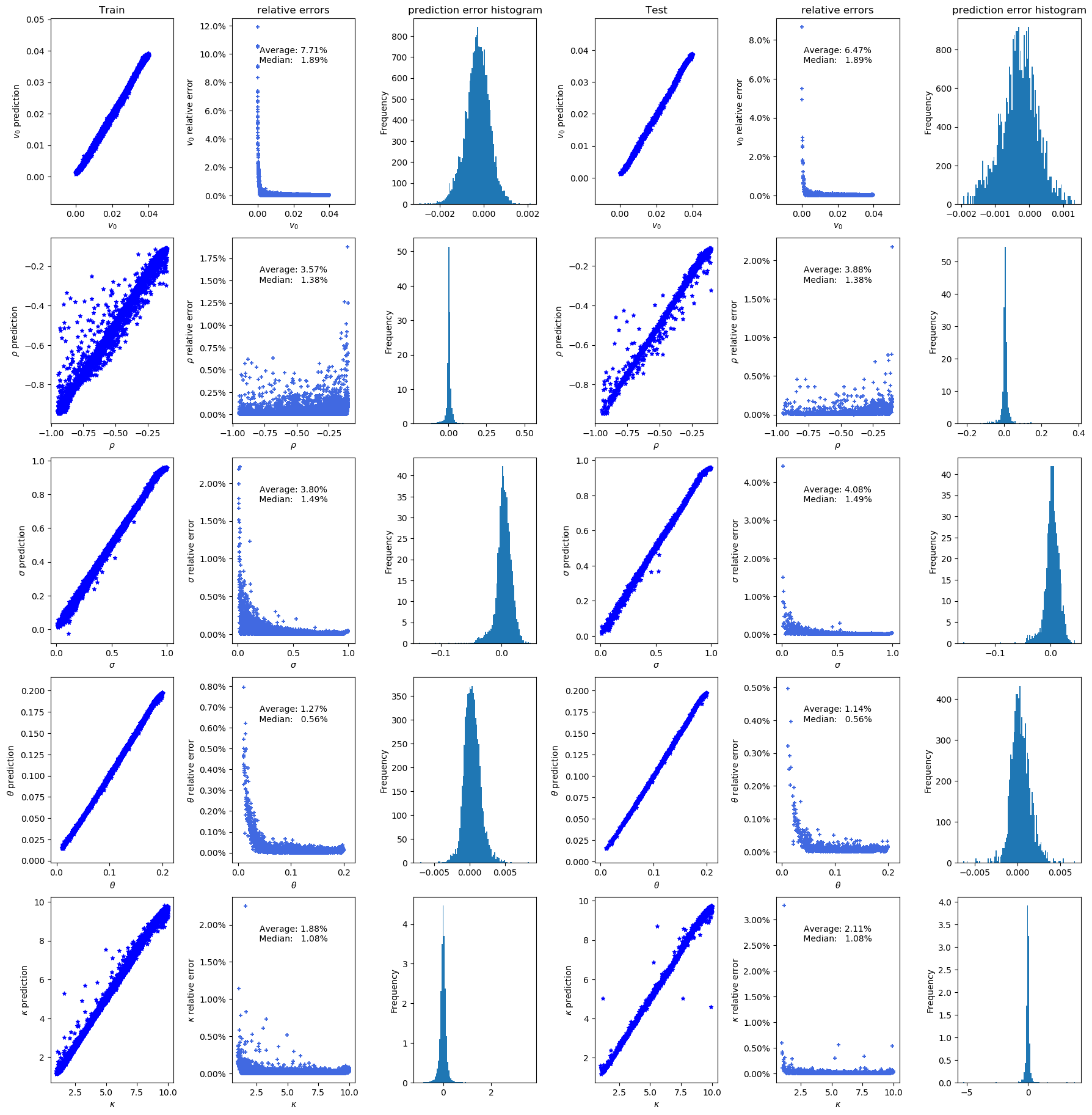}
    \caption{CNN prediction errors.}
    \label{fig:C_err}
\end{figure}

Figure~\ref{fig:F_err} and Figure~\ref{fig:C_err} display the errors between the predicted Heston model parameters and the labels for the FCNN and the CNN. The fact that the performance of the NNs is good for both the training and testing data implies that the NN models are able to perform well when predicting unseen data. Notice that only the predictions for $\rho$ are slightly worse as compared to the other four model parameters.

As can be seen in the figures, the relative errors obtained by the FCNN are significantly smaller. Although it has been found that, in general, CNNs outperform FCNNs, especially in tasks such as image recognition, our results indicate that this is not the case when calibrating the Heston model. The CNN applies $3\times 3$ kernel filters and maxpooling to the training data which, to some extent,  causes information loss, while the FCNN avoids this problem. For image recognition this is not an issue since the main features of the image are of most importance to the NN, whereas the Heston model calibration is a regression problem that is found to be overly sensitive to the information loss caused by this type of layer operations. Note that in image recognition the goal is often to simply label an image, whereas to calibrate the Heston model the exact values for the parameter outputs are required instead.

Another main handicap when trying to learn the Heston model calibration using a NN is the model parameter identifiability due to the nature of the volatility matrices. As~\cite{roeder2020volatility} mentioned, there exist Heston parameterisations which differ significantly on the values of the Heston parameters but correspond to similar implied volatility surfaces. This explains why in the case of both NN structures, although most of the predictions are quite accurate and relatively close to the $0.0\%$ error mark, several outliers that are substantially far from the rest of the data points can be found. This can also be appreciated when comparing the median and average errors.

\section{Discussion of interpretability results}
\label{sec:discussion}

In this section we will discuss the interpretability results obtained by using different methods. We wish to understand if local or global methods are best to interpret the NNs, and whether they could be used for other more complex and less understood models beyond the Heston model.

\subsection{Local interpretability results: local surrogate methods}
\label{sec:local}


Local surrogate methods (or models) are interpretable methods used to explain single predictions of black box machine learning models. From the following results we find that in general local interpretability does not align with our intuition behind the Heston model.

\subsubsection{LIME}

LIME (Local Interpretable Model-agnostic Explanations) is primarily designed to explain classifiers and NNs performing image recognition. In our case we treat the NN as a regressor and we first perform a local approximation around an individual prediction using an explanation model (here a linear model is used). In particular, the Huber Regressor is chosen for its robustness against outliers.


\begin{figure}
	\centering
    \includegraphics[width=0.9\textwidth,keepaspectratio]{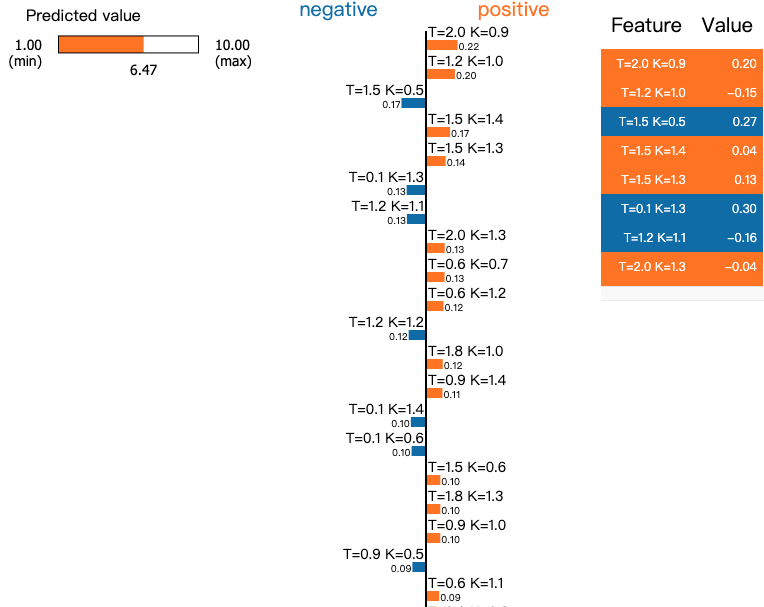}
	\caption{FCNN LIME attributions of $\kappa$ at 0th observation.}
	\label{fig:F_lime_kappa}
\end{figure}

\begin{figure}
	\centering
    \includegraphics[width=0.9\textwidth,keepaspectratio]{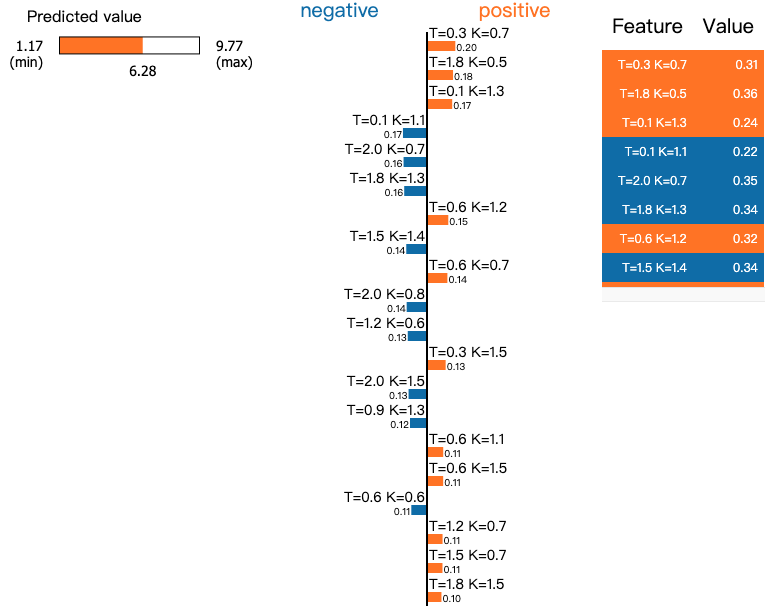}
	\caption{CNN LIME attributions of $\kappa$ at 0th observation.}
	\label{fig:C_lime_kappa}
\end{figure}

We consider as an example the first $\kappa$ in the test set. As shown in Figure~\ref{fig:C_lime_kappa}, over 1500 predictions ranging from 1.17 to 9.77, the first predicted value for the CNN is 6.28. For this particular prediction, the most important feature is $(T=0.3, K=0.7)$ close to the at-the-money (ATM) level with a short maturity, followed by $(T=1.8, K=0.5)$ with a very long maturity far away from the ATM level. They both influence the predicted value positively, since they are coloured in orange. Blue coloured components, on the other hand, have a negative contribution to the model output. The feature-value table specifies the input value of each feature which is also coloured either in orange or blue.

As for the FCNN in Figure~\ref{fig:F_lime_kappa}, the predicted value is 6.47 which is quite close to the value predicted by the CNN. We notice that $\kappa$ is generated from 1.0 to 10.0 according to Table~\ref{tab:para_bounds}, which indicates that the FCNN seems to cover the entire range of the model parameter. In the case of the FCNN, the topmost features concentrate around long term maturities such as $T=2.0$, $T=1.2$ and $T=1.5$ and the top two strikes are close to the ATM levels.



\begin{figure}
	\centering
    \includegraphics[width=.50\textwidth,keepaspectratio]{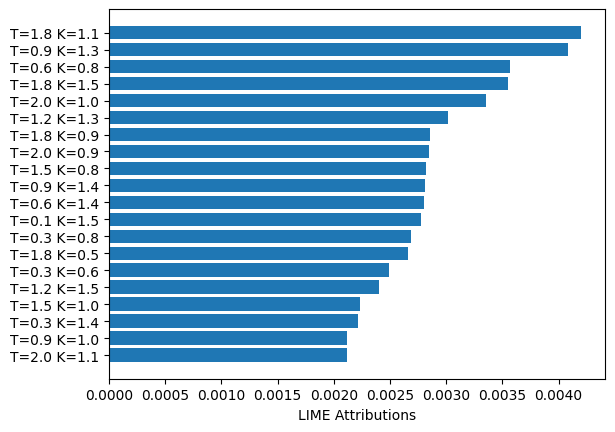}
    \hspace*{1em}
	\includegraphics[width=.40\textwidth,keepaspectratio]{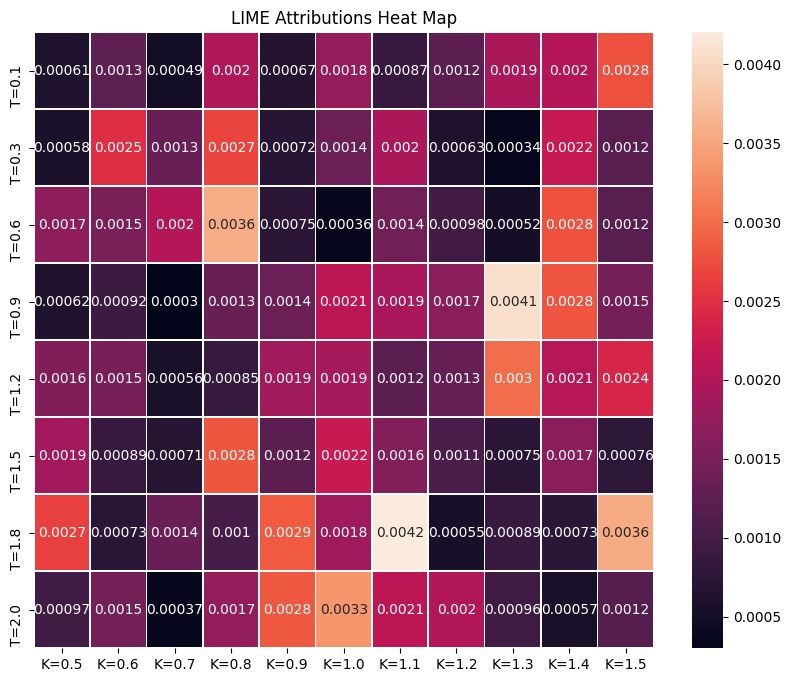}
	\caption{FCNN LIME attributions and heat map.}
	\label{fig:F_lime}
\end{figure}
\begin{figure}
	\centering
    \includegraphics[width=.50\textwidth,keepaspectratio]{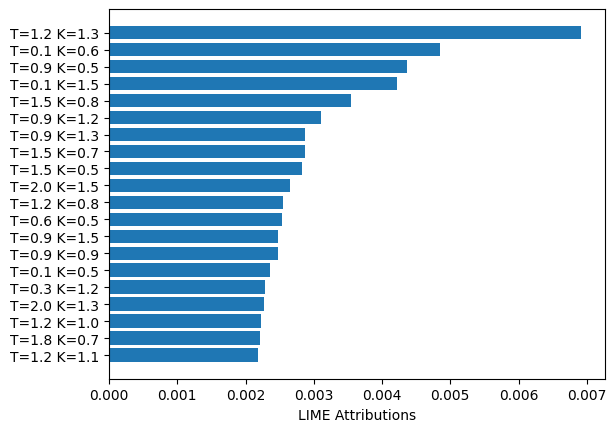}
    \hspace*{1em}
	\includegraphics[width=.40\textwidth,keepaspectratio]{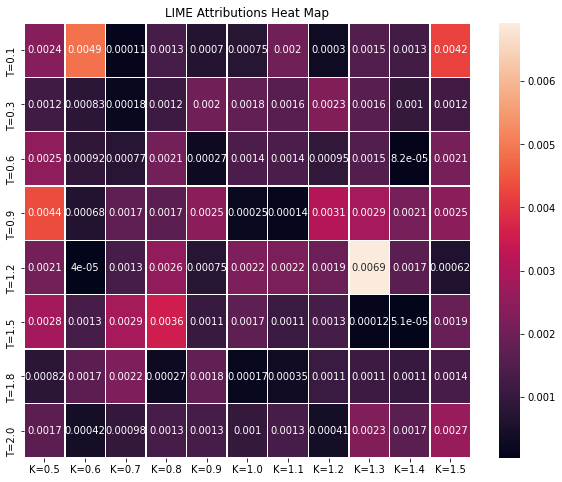}
	\caption{CNN LIME attributions and heat map.}
	\label{fig:C_lime}
\end{figure}

We continue by looking at the overall impact of the input entries on the model output by taking the absolute value of each feature importance and averaging them over the 1500 predictions. From these, we obtain Figure~\ref{fig:F_lime} and Figure~\ref{fig:C_lime}. The light colours in the heat maps indicate high attribution values for the input and dark colours low attribution values. The most influential volatilities seem to be randomly located in these two heat maps and do not show a particular pattern.

We would like to highlight that the attribution results shown here are just one of the possible results that LIME might find. LIME may predict different feature attributions for the same instance since it uses an optimization algorithm with a random seed that can lead to different local minima and give different results every time it is run. This suggests that the mapping function for the Heston model is highly non-linear and may contain multiple local minima, which makes LIME not ideal for this application.

\subsubsection{DeepLIFT}

\texttt{DeepExplain} in Python provides the implementation for re-scaled DeepLIFT, in which a modified chain rule is applied as discussed in~\cite{ancona2018towards}. Attributions are computed through the test data set, and the baseline $\bar{x}$ here is chosen by default to be a zero array with the same size as the input. These attributions have shape (1500, 8, 11) for the CNN and (1500, 88) for the FCNN. To draw the feature importance, the attributions of the CNN should be reshaped to (1500, 88). Again, we take the absolute values of each feature importance and we average them over the 1500 predictions.

\begin{figure}
	\centering
    \includegraphics[width=.50\textwidth,keepaspectratio]{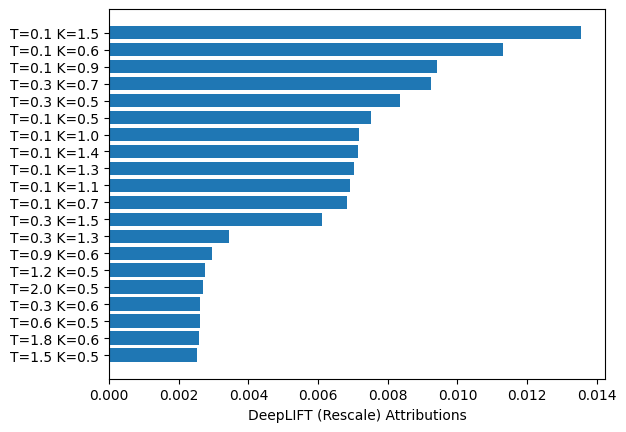}
    \hspace*{1em}
	\includegraphics[width=.40\textwidth,keepaspectratio]{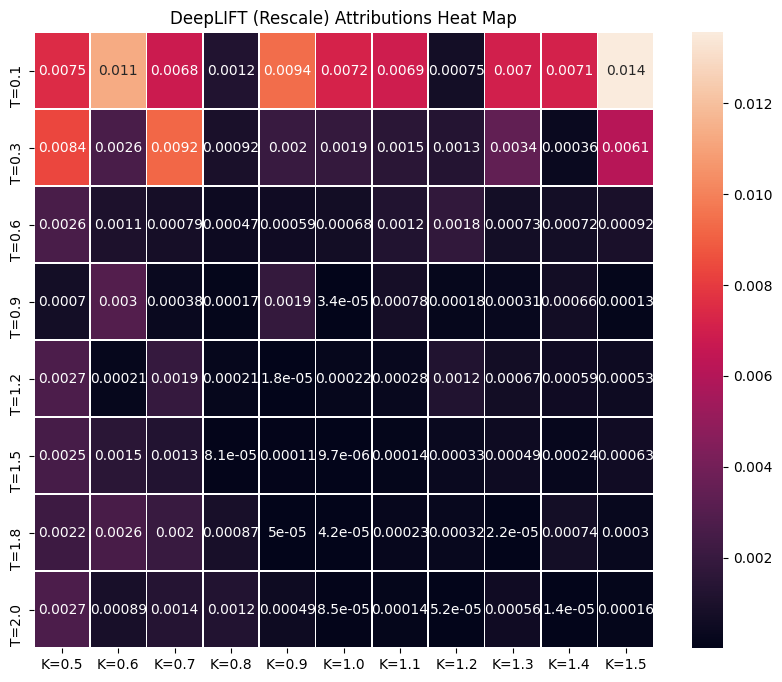}
	\caption{FCNN DeepLIFT attributions and heat map.}
	\label{fig:F_deeplift}
\end{figure}

\begin{figure}
	\centering
    \includegraphics[width=.50\textwidth,keepaspectratio]{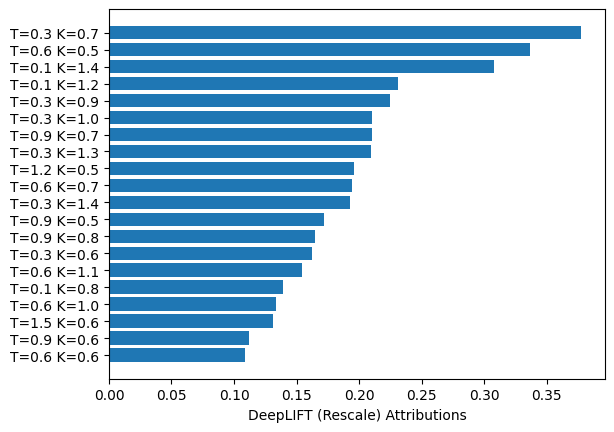}
    \hspace*{1em}
	\includegraphics[width=.40\textwidth,keepaspectratio]{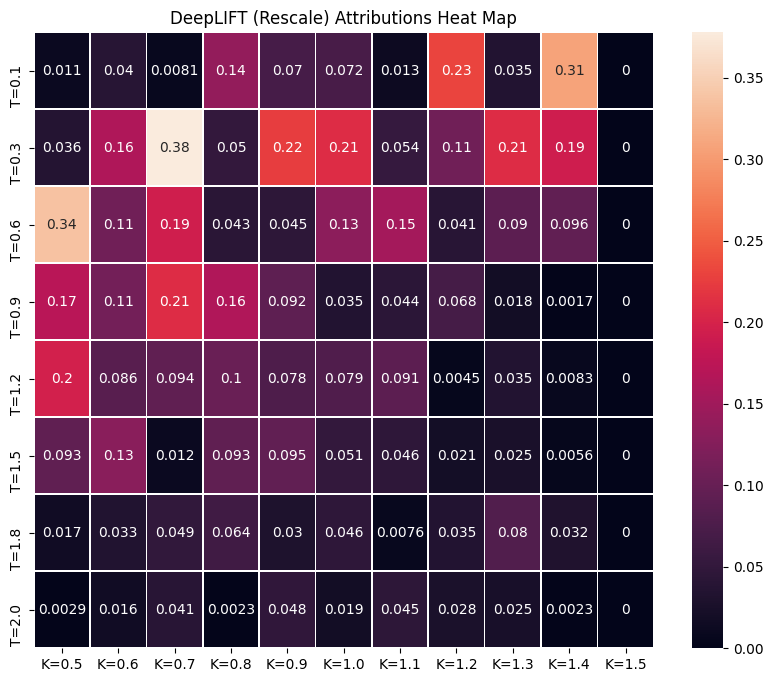}
	\caption{CNN DeepLIFT attributions and heat map.}
	\label{fig:C_deeplift}
\end{figure}

Figure~\ref{fig:F_deeplift} and Figure~\ref{fig:C_deeplift} present the overall impact on the model output $\psi$. It can be seen that the lightest colours appear on both sides of the strike range, at the wings, and in the positions within the $T=1.2$ borderline; while in the case of the FCNN, we can see that the lightest colours are located around short maturities, $T=0.1$ and $T=0.3$. However, its topmost features are located at extreme strikes, $K=1.5$ and $K=0.6$.

\subsubsection{LRP}

The Layer-wise Relevance Propagation (LRP) algorithm aims at attributing relevance to individual input nodes to trace back contributions to the output, $\psi$, layer by layer. Although several version of LRP exist, we focus on the $\epsilon$-LRP algorithm which makes use of the $\epsilon$-rule as described by~\cite{bach2015pixel}. $\epsilon$ must be non-zero and in our case, we set it to the default 0.0001 value. LRP can be implemented in \texttt{DeepExplain} too.

\begin{figure}
	\centering
    \includegraphics[width=.50\textwidth,keepaspectratio]{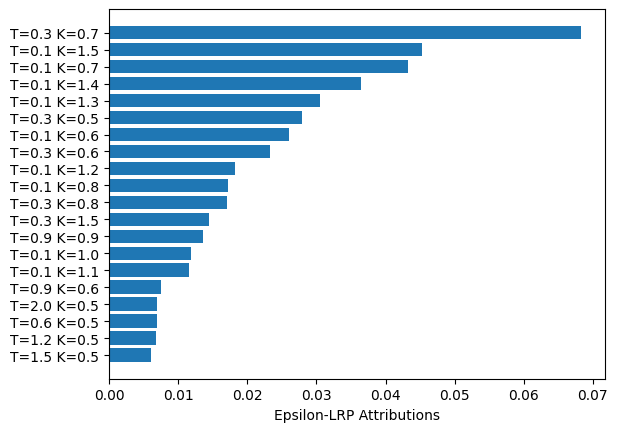}
    \hspace*{1em}
	\includegraphics[width=.40\textwidth,keepaspectratio]{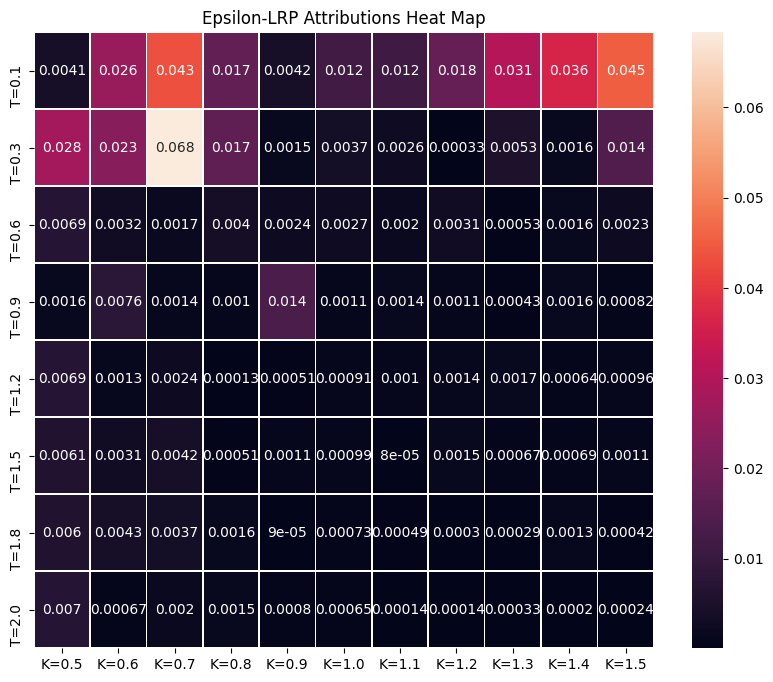}
	\caption{FCNN LRP attributions and heat map.}
	\label{fig:F_LRP}
\end{figure}

\begin{figure}
	\centering
    \includegraphics[width=.50\textwidth,keepaspectratio]{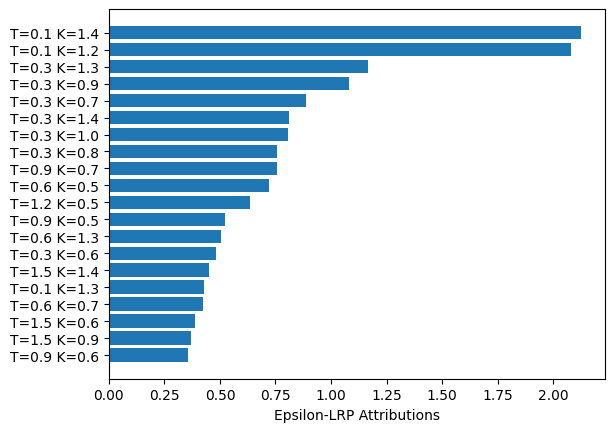}
    \hspace*{1em}
	\includegraphics[width=.40\textwidth,keepaspectratio]{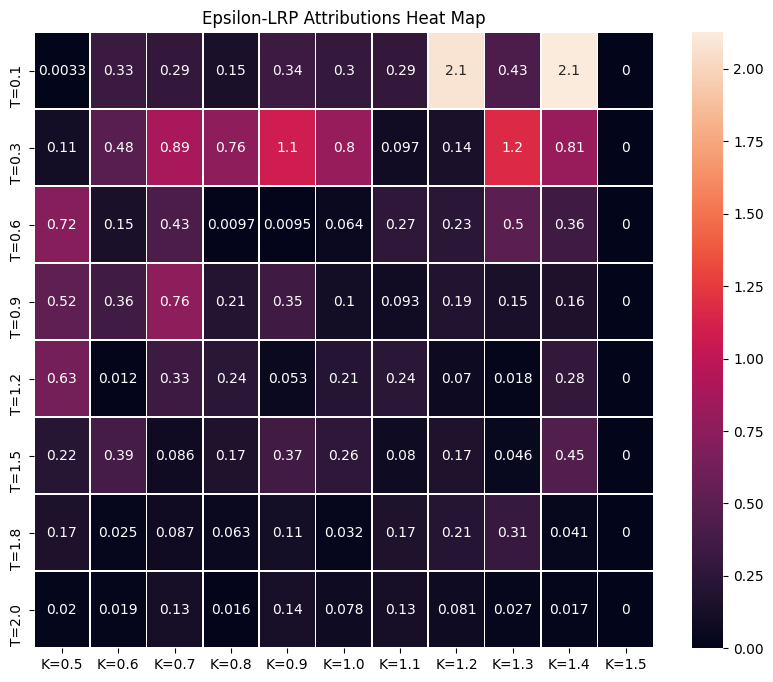}
	\caption{CNN LRP attributions and heat map.}
	\label{fig:C_LRP}
\end{figure}

Using LRP we find that the most important features for the FCNN are the ones with strikes at both the left and the right wing: $K=0.7$, $K=1.5$, and $K=1,4$, as shown in Figure~\ref{fig:F_LRP}. On the other hand, from Figure~\ref{fig:C_LRP}, we see that in the case of the CNN, the top features are at large strikes, $K=1.4$, $K=1.2$ and $K=1.3$. Overall, the most relevant features for both the CNN and the FCNN are located at volatilities of short maturities and extreme strikes.

\subsection{Global interpretability results: Shapley values}
\label{sec:global}

We recall that the size of the test subset is $1{,}500$ so the Shapley values have dimensions (1500, 88) for the FCNN since the input is flattened in this case and (1500, 8, 11) for the CNN. We measure the feature importance of the model input by taking the mean of the absolute Shapley values, which is representative of the average impact on the model output. We present here a selection of summary plots and we also refer the reader to Appendix~\ref{sec:force} for a more detailed analysis performed by visualizing the Shapley values as ``forces'' as described in~\cite{molnar2019}.

\begin{figure}
\centering
\includegraphics[width=0.95\textwidth,keepaspectratio]{FNN_shap_seperate.png}
\caption{FCNN SHAP values and feature importance of each model parameter.}
\label{fig:FNN_SHAP_seperate}
\end{figure}

\begin{figure}
\centering
\includegraphics[width=0.95\textwidth,keepaspectratio]{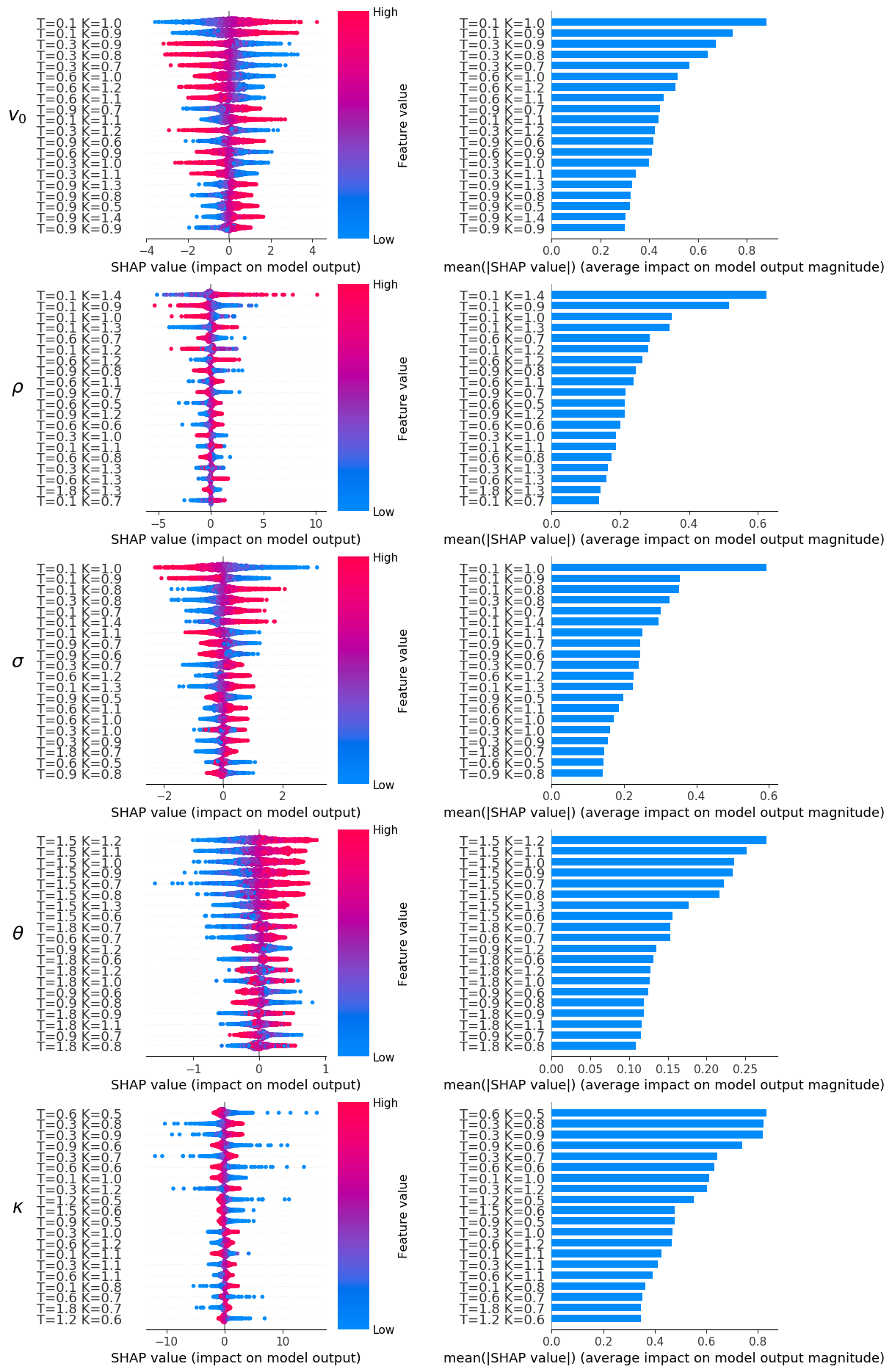}
\caption{CNN SHAP values and feature importance of each model parameter.}
\label{fig:CNN_SHAP_seperate}
\end{figure}

The SHAP summary plots shown on the left of Figure~\ref{fig:FNN_SHAP_seperate} and Figure~\ref{fig:CNN_SHAP_seperate} combine both feature importance and effects. Each point drawn on the summary plot corresponds to a Shapley value for a feature and an instance. The position on the x-axis is determined by the Shapley value and the position on the y-axis by the feature. For the Heston model calibration, each feature is an entry of the input volatility matrix for a given maturity and strike. The features are ordered from most to least important. On the other hand, the plots shown on the right of Figure~\ref{fig:FNN_SHAP_seperate} and Figure~\ref{fig:CNN_SHAP_seperate} are called the SHAP Feature Importance plots. We display one of these plots for each of the Heston model parameters. The aim of the plots is to highlight the importance of features with large absolute Shapley values. We average the absolute Shapley values per feature across the data to obtain the global importance. Once again, the features are sorted from most to least important, see~\cite{molnar2019}. Each row of the aforementioned plots represents one model parameter of the calibrated Heston model: $v_0, \rho, \sigma, \theta$ and $\kappa$ from top to bottom. Although all Shapley values were calculated, here we only show the top 20 most important for practical reasons.

From these figures we can see that the features which affect $v_0$ the most are the plain-vanilla options close to the ATM level. Yet, we can see a different behaviour between the FCNN and CNN architectures. In the case of the FCNN, we notice that only short maturities are found amongst the most relevant features, while for the CNN, longer maturities are present. Since $v_0$ represents the starting value of the (squared) volatility process, it should be linked to near-ATM plain-vanilla options at short maturities, a correspondence supported only by the FCNN architecture. Moreover, we recall that the CNN obtains its worse calibration results for $v_0$ with an average error of $6.47\%$ for the test data, while the FCNN only produces an error of $1.07\%$ for this parameter, as shown in Figures~\ref{fig:F_err} and~\ref{fig:C_err}. We continue by looking at the correlation $\rho$ and the volatility of volatility $\sigma$. These parameters are linked to the asymmetry and the convexity of the smile, respectively. They should be linked to plain-vanilla options with strikes far from the ATM level, mainly at shorter maturities, since at higher maturities the smile flattens. Indeed, we find this correspondence well described by the FCNN architecture, but we are not fully satisfied by the CNN architecture which behaves well for $\rho$, but for $\sigma$ it seems to give importance mainly to ATM options. Then, we consider the speed of mean reversion $\kappa$ which is related to the flattening of the smile for longer maturities. We find that for both architectures the plain-vanilla options with most importance are close to the wings and with both shorter and longer maturities. Finally, we consider the mean-reversion level $\theta$, which is expected to be mostly influenced by the long-term behaviour of the data. In this case, both NN architectures show similar results and agree with our intuition.

\begin{figure}
	\centering
    \includegraphics[width=.50\textwidth,keepaspectratio]{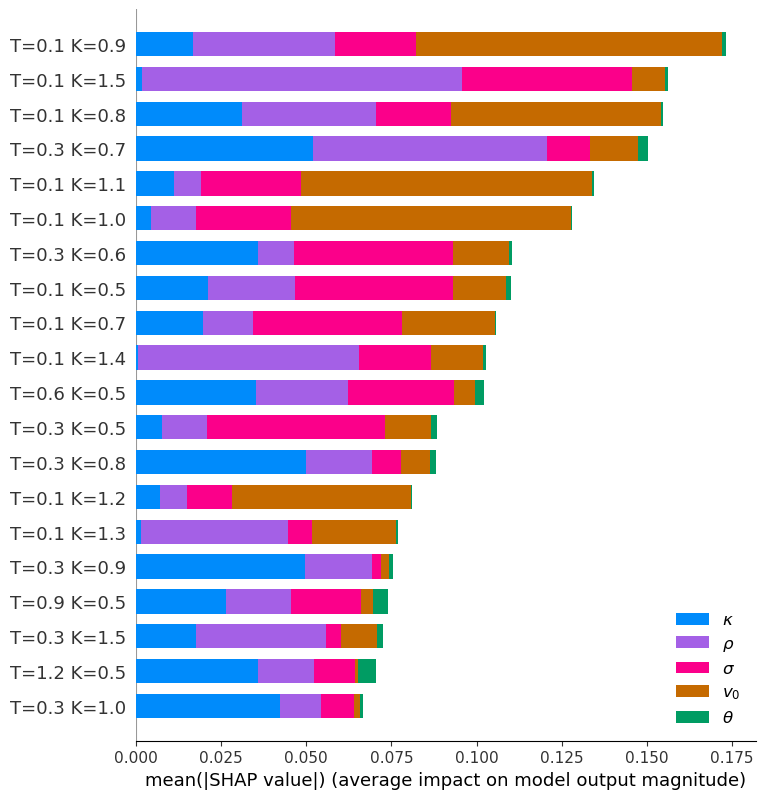}
    \hspace*{1em}
	\includegraphics[width=.40\textwidth,keepaspectratio]{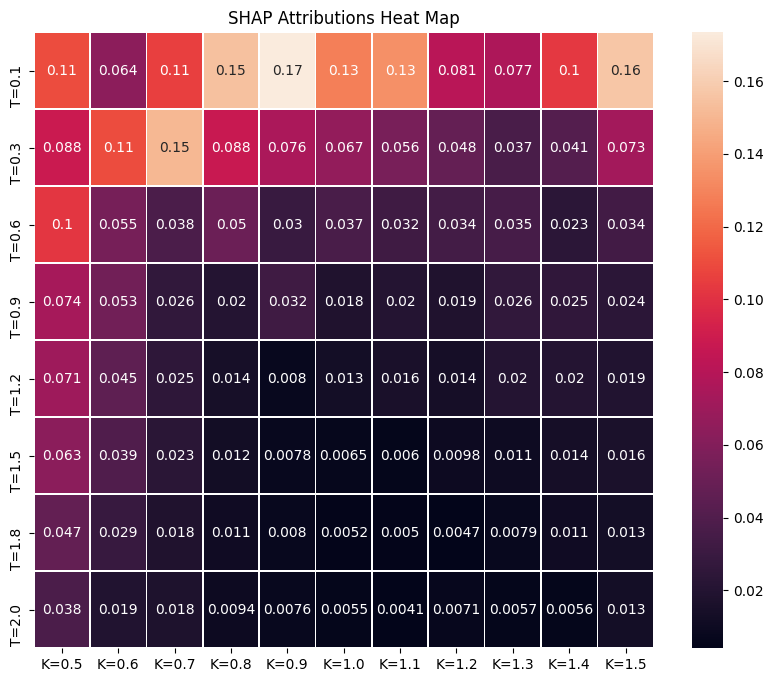}
	\caption{FCNN SHAP values overall feature importance and heat map.}
	\label{fig:SHAP_allfcnn}
\end{figure}
\begin{figure}
	\centering
    \includegraphics[width=.50\textwidth,keepaspectratio]{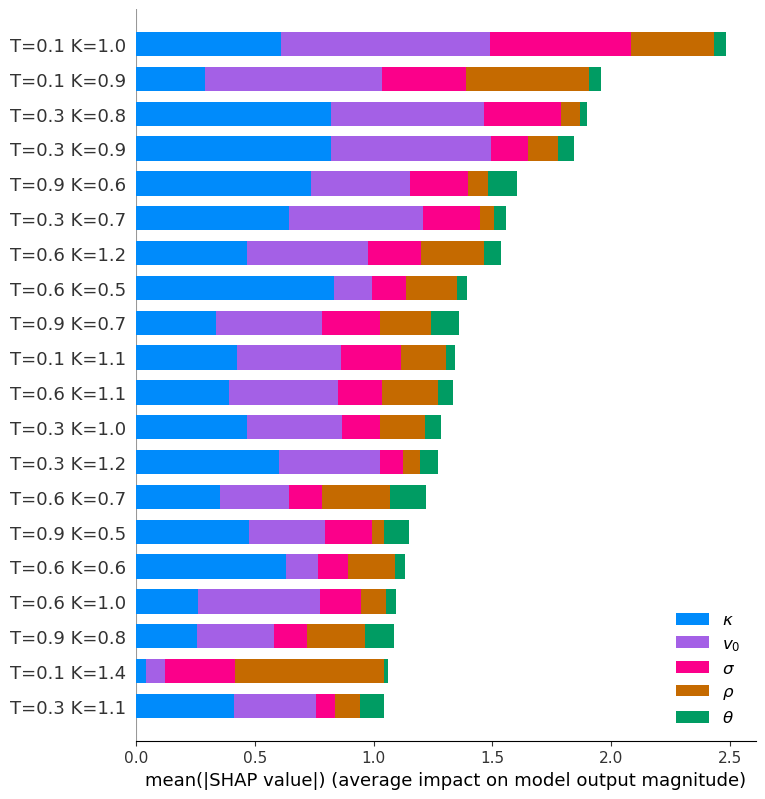}
    \hspace*{1em}
	\includegraphics[width=.40\textwidth,keepaspectratio]{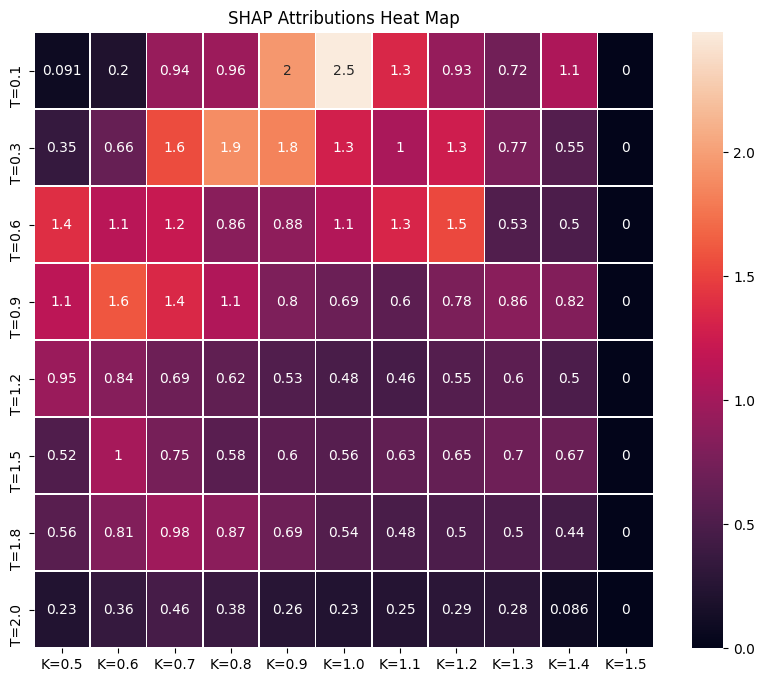}
	\caption{CNN SHAP values overall feature importance and heat map.}
	\label{fig:SHAP_allcnn}
\end{figure}

In
Figure~\ref{fig:SHAP_allfcnn} and Figure~\ref{fig:SHAP_allcnn}, we summarise the feature importance for the overall model output $\psi$. It is equivalent to summing the previous Shapley values for the set of parameters $v_0, \rho, \sigma, \theta$ and $\kappa$. It is interesting to note that the top 10 most important features of the FCNN seem to be consistently at shorter maturities and not necessarily at ATM levels, whereas in the case of the CNN, they are all close to ATM levels, but the maturity fluctuates substantially.

We obtain two important results from this analysis. First, we can view Shapley values as a practical tool to discriminate among NN architectures to select the ones that better match the model behaviour by establishing clear correspondences between calibrated options and model parameters. Second, we can use this tool to investigate a model's behaviour when we are confident of the NN architecture but we lack a clear model interpretation.

\section{Conclusion and further developments}
\label{sec:conclusion}

In this paper we have first explored the volatility smile to model parameters relationship (``calibration'') in the Heston model, using deep learning, and compared different networks used in the literature. We conclude that Fully-Connected Neural Networks (FCNNs) outperform Convolutional Neural Networks (CNNs) in pricing-model calibration, since they require less trainable parameters and obtain better results. Next, the bulk of our analysis has focused on comparing local and global interpretability methods and evaluating their ability to give meaningful insights into the pricing-model calibration mechanism using NNs. We have used the Heston model as our benchmark because it is a well understood model both theoretically and in practice. From our analysis we find that global interpretability methods seem to be most reliable and substantially align with the common intuition behind the Heston model. They may also help in choosing the most convenient type of NN, favouring FCNNs over CNNs in our case. 
Hence, global interpretability methods are encouraged.

In future work, the Feller condition could be imposed not only on the training data but also on the model output, local interpretability could be extended by trying to explain, for each prediction, the attributions for the individual model parameters $v_0$, $\rho$, $\sigma$, $\theta$, and $\kappa$ instead of the output $\psi$ as a whole, and the calibration could be performed on real market data.

\bibliography{biblio}
\bibliographystyle{apalike}

\appendix

\section{Details on local interpretability methods}
\label{sec:lmdetail}

Here, we describe in greater detail the local interpretability methods, and we highlight the particular choices we adopted in their implementation.

\subsection{Local interpretable model-agnostic explanations (LIME)}

Local interpretable model-agnostic explanations (LIME) is a technique that explains predictions of any classifier or regressor by learning an interpretable model locally around the prediction. It was first presented in~\cite{lime}. LIME is an additive feature attribution method since it uses a local linear explanation model that is consistent with equation~\eqref{eq:AFA}. Here, the simplified inputs $x'$ are considered as interpretable inputs. The mapping function $h_x$ maps a binary vector of interpretable inputs $x'$ into the original input space, that is, $x= h_x(x')$. Different input spaces use different mapping functions. Mathematically, LIME aims at minimising the following objective function:
\begin{equation}\label{eq:lime_obj_func}
  \xi=\underset{g \in \mathcal{G}}{\arg \min} \, L\left(f, g, \pi_{x^{\prime}}\right)+\Omega(g)
\end{equation}%
where $g$ is the explanation model for instance $x'$ that minimise the loss function $L$, and $\mathcal{G}$ is a family of possible explanations. $g$ measures how close the explanation is to the prediction of the original model. $\Omega$ penalises the complexity of $g$ and thus, $\Omega(g)$ is the model complexity in the minimization. The local kernel $\pi_{x^{\prime}}$ are the weights in the loss $L$ over a set of samples in the simplified input space, which measures how large the neighborhood around instance $x'$ is. The explanation model $g$ here satisfies equation~\eqref{eq:AFA} and the loss $L$ is taken to be a squared loss. Thus, we can solve the objective function~\eqref{eq:lime_obj_func} by using penalised linear regression.

\subsection{Deep-learning important features (DeepLIFT)}

Deep-learning important features (DeepLIFT) is a method that explains the prediction of deep learning NNs recursively. It was first presented in~\cite{Shrikumar2016NotJA,Shrikumar2017LearningIF}. For each input $x_i$, the attribution $C_{\Delta x_i \Delta_y}$ is computed by setting the effect of that input $x_i$ to a reference value which is opposed to its original value. In this case, the mapping function $h_x$ converts the binary vector $x'$ into the original input space: $x=h_x(x')$ with $1$ in $x'$ meaning that the original values are taken and $0$ meaning that the reference value is taken instead. The authors specify the reference value as a typical uninformative background value for the feature.

Also, a summation-to-delta property should be satisfied in DeepLIFT: 
\begin{equation*}
    \sum_{i=1}^n  C_{\Delta x_i \Delta t} = \Delta t
\end{equation*}%
where $t=f(x)$ is the model output, and $\Delta t = t-t^0 = f(x) -f(a)$ is the difference from the reference input value $a$. If we set $\phi_i = C_{\Delta x_i \Delta t}$, then the explanation model has the form of equation~\eqref{eq:AFA}.

Furthermore, in our context DeepLIFT is implemented with a modified chain rule proposed by~\cite{ancona2018towards}. Mathematically, we define $z_{ij} = w_{ji}^{(l+1,l)} x_i^{(l)}$ to be the weighted activation of a neuron $i$ of layer $l$ onto neuron $j$ in the next layer, and $b_j$ the additive bias of unit $j$. Each attribution of unit $i$ indicates the relative effect of the unit activated at the original input $x$ compared to the activation at some reference input $\bar{x}$ (baseline). Reference values $\bar{z}_{ji}$ for all hidden units are obtained by running a forward pass through the NN with input $\bar{x}$, and recording the activation of each unit. The baseline is to be determined by the user and is often chosen to be zero. Therefore, let $\phi_i^c(x) = r_i^{(l)}$ be the attributions at the input layer for neuron $c$, then:

\begin{equation*}
    r_{i}^{(L)}=\left\{\begin{array}{ll}S_{i}(x)-S_{i}(\bar{x}) & \text { if unit } i \text { is the target unit of interest } \\ 0 & \text { otherwise }\end{array}\right.
\end{equation*}%
where the algorithm starts with the output layer $L$. Finally, the Rescale Rule is defined as 
\begin{equation*}
    r_{i}^{(l)}=\sum_{j} \frac{z_{j i}-\bar{z}_{j i}}{\sum_{i^{\prime}} z_{j i}-\sum_{i^{\prime}} \bar{z}_{j i}} r_{j}^{(l+1)}.
\end{equation*}
The modified chain rule is adopted and the original Reveal-Cancel rule from~\cite{Shrikumar2017LearningIF} is not considered for implementation purposes.

\subsection{Layer-wise relevance propagation (LRP)}

Layer-wise relevance propagation (LRP) is closely related to DeepLIFT. It was first presented in~\cite{bach2015pixel}. If we fix the reference activations of all neurons to be 0 in DeepLIFT, it becomes LRP. Hence, for the mapping function $x=h_x(x')$, 1 in the binary vector $x'$ represents that the original value of an input is taken, and 0 means that 0 values are taken. It also satisfies the additive feature attribution method structure given in equation~\eqref{eq:AFA}.

\section{SHAP force plots and clustering values}
\label{sec:force}
 
Shapley values can be visualised as ``forces''. Each feature value can be thought of as a force that either contributes to increasing or decreasing the prediction. The prediction starts from the average of all predictions, which is taken as the baseline, which may also be called the base value. Each Shapley value is then represented as an arrow in the plot, that either pushes to increase or decreases the model output. Positive values that push to increase the prediction are represented in red and negative values that push to decrease the prediction in blue. Note that here ``positive'' and ``negative'' do not make reference to the Shapley value being ``good'' or ``bad'' for the prediction, but rather moving the prediction to the right or to the left with respect to the base value. The red and blue colour features are stacked together on the edges of the force plot to show their values on hover and how much those features influence the final output value. The wider the Shapley value represented by the force plot, the bigger the influence it has on the final prediction. All the forces balance each other out at the actual prediction as described by~\cite{molnar2019}. This type of representations may be helpful to get a better intuition of the results obtained in Figure~\ref{fig:FNN_SHAP_seperate} and Figure~\ref{fig:CNN_SHAP_seperate}.

\begin{figure}
	\centering
    \includegraphics[width=0.95\textwidth,keepaspectratio]{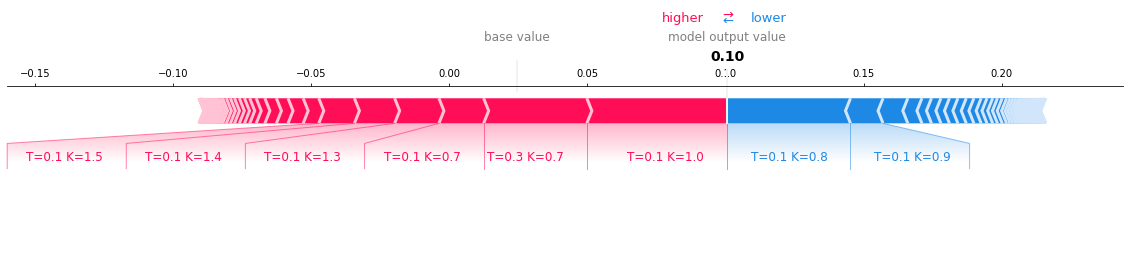}
	\caption{FCNN force plot for $v_0$ (First Observation)}
	\label{fig:SHAP_force11}
\end{figure}

\begin{figure}
	\centering
    \includegraphics[width=0.95\textwidth,keepaspectratio]{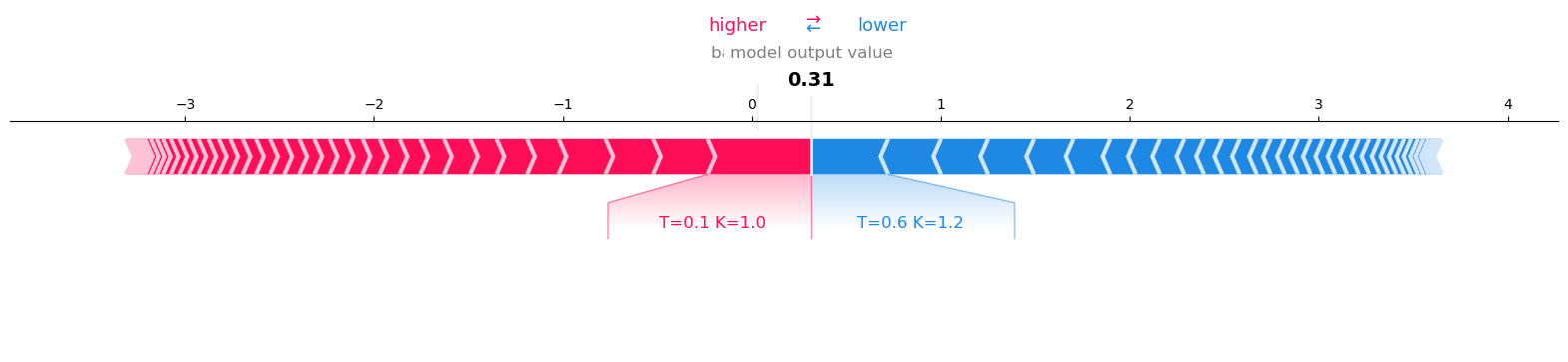}
	\caption{CNN force plot for $v_0$ (First Observation)}
	\label{fig:SHAP_force22}
\end{figure}

Figure~\ref{fig:SHAP_force11} and Figure~\ref{fig:SHAP_force22} display two sample force plots for the FCNN and the CNN, respectively. These force plots are generated using the first observation out of the 1500 samples in the test set to predict the model output $v_0$. Let's explain how to interpret Figure~\ref{fig:SHAP_force11}. The base value lies between 0.5 and 0.55 and it is indicated by a grey vertical line. On the other hand, the model output value is 0.61, which is highlighted in bold and its position on the plot is indicated by a vertical grey line too. As expected the base value and model output do not coincide, although they are relatively close. Next, the most relevant Shapley values are highlighted at the bottom of each force plot. We can clearly appreciate that the red or ``positive'' Shapley values outweigh the blue or ``negative'' Shapley values, therefore the model output value is to the right of the base value. A similar behaviour can be observed in Figure~\ref{fig:SHAP_force22}. The red Shapley values also outweigh the blue ones and the model output is moved to the right with respect to the base value. 

Mind the scale in Figure~\ref{fig:SHAP_force22}, since there is actually a bigger offset between the model output value and the base value in Figure~\ref{fig:SHAP_force22} than in Figure~\ref{fig:SHAP_force11}. The difference in scale between Figure~\ref{fig:SHAP_force11} and Figure~\ref{fig:SHAP_force22} can be attributed to the fact that the FCNN mainly focuses its attention on a few inputs, hence the Shapley values for the least relevant inputs have a small influence in the final model output value, its Shapley values are represented by really short arrows and this results in a more compact force plot. In contrast, in Figure~\ref{fig:SHAP_force22}, the importance of each Shapley is relatively comparable for all inputs, this can be attributed to the fact that the CNN applies convolution and maxpooling to the original input. These operations have an ``averaging'' effect on the input information and importance, which results in the NN struggling to focus its attention on the most relevant features (this also makes the CNN perform worse), therefore the Shapley values on the force plot are represented by longer arrows and the overall force plot is wider for the CNN.

\begin{figure}
	\centering
	\includegraphics[width=0.95\textwidth,keepaspectratio]{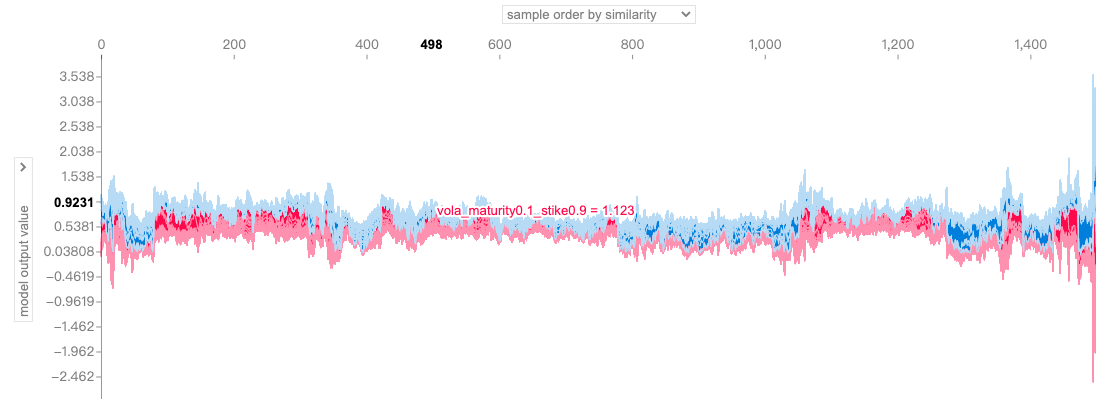}
	\caption{FCNN SHAP clustering for $v_0$  (All Data)}
	\label{fig:SHAP_force_allfcnn}
\end{figure}

\begin{figure}
	\centering
    \includegraphics[width=0.95\textwidth,keepaspectratio]{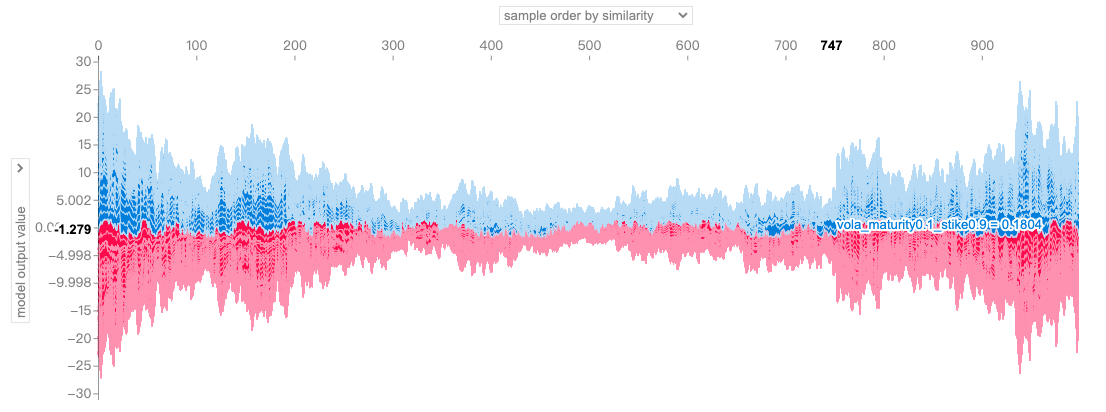}
	\caption{CNN SHAP clustering for $v_0$  (All Data)}
	\label{fig:SHAP_force_allcnn}
\end{figure}

Figure~\ref{fig:SHAP_force_allfcnn} and Figure~\ref{fig:SHAP_force_allcnn} are known as SHAP clustering plots. They corresponds to creating force plots for all 1500 observations and stacking them together vertically in a single plot. The x-axis in Figure~\ref{fig:SHAP_force11} is equivalent to the y-axis in Figure~\ref{fig:SHAP_force_allfcnn}. The x-axis in Figure~\ref{fig:SHAP_force_allfcnn} is the observation number. We can clearly see that as previously mentioned, the force plots for the CNN are notably wider. Force plots of the other model parameters $ \rho, \sigma, \theta$ and $\kappa$ can be visualised with the same method.

\section{Additional results using a larger data set}
\label{largerdataset}

\begin{figure}
    \centering
    \includegraphics[width=1.0\textwidth,keepaspectratio]{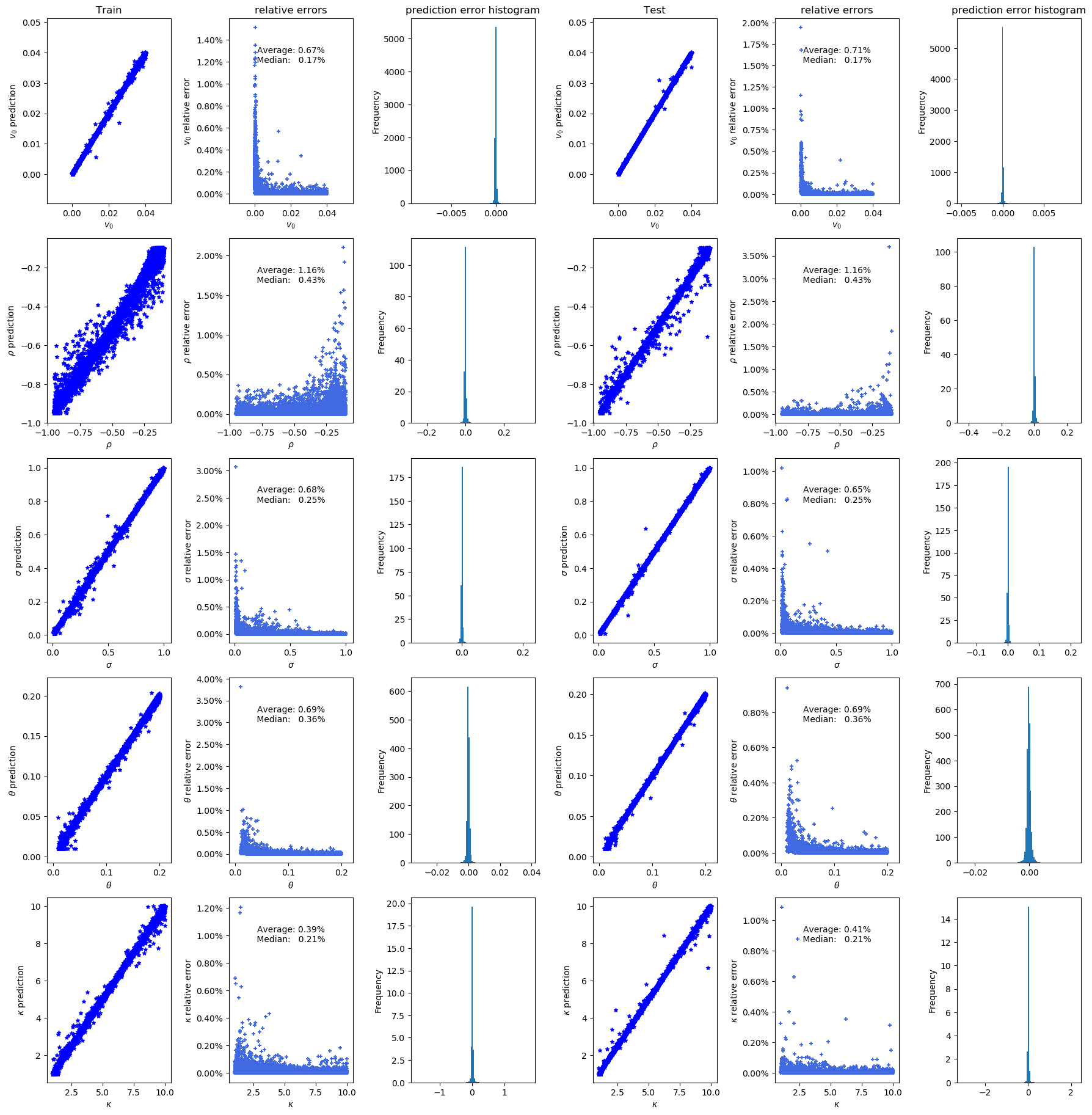}
    \caption{FCNN prediction errors for large data set.}
    \label{fig:F_histlarge}
\end{figure}

\begin{figure}
    \centering
    \includegraphics[width=1.0\textwidth,keepaspectratio]{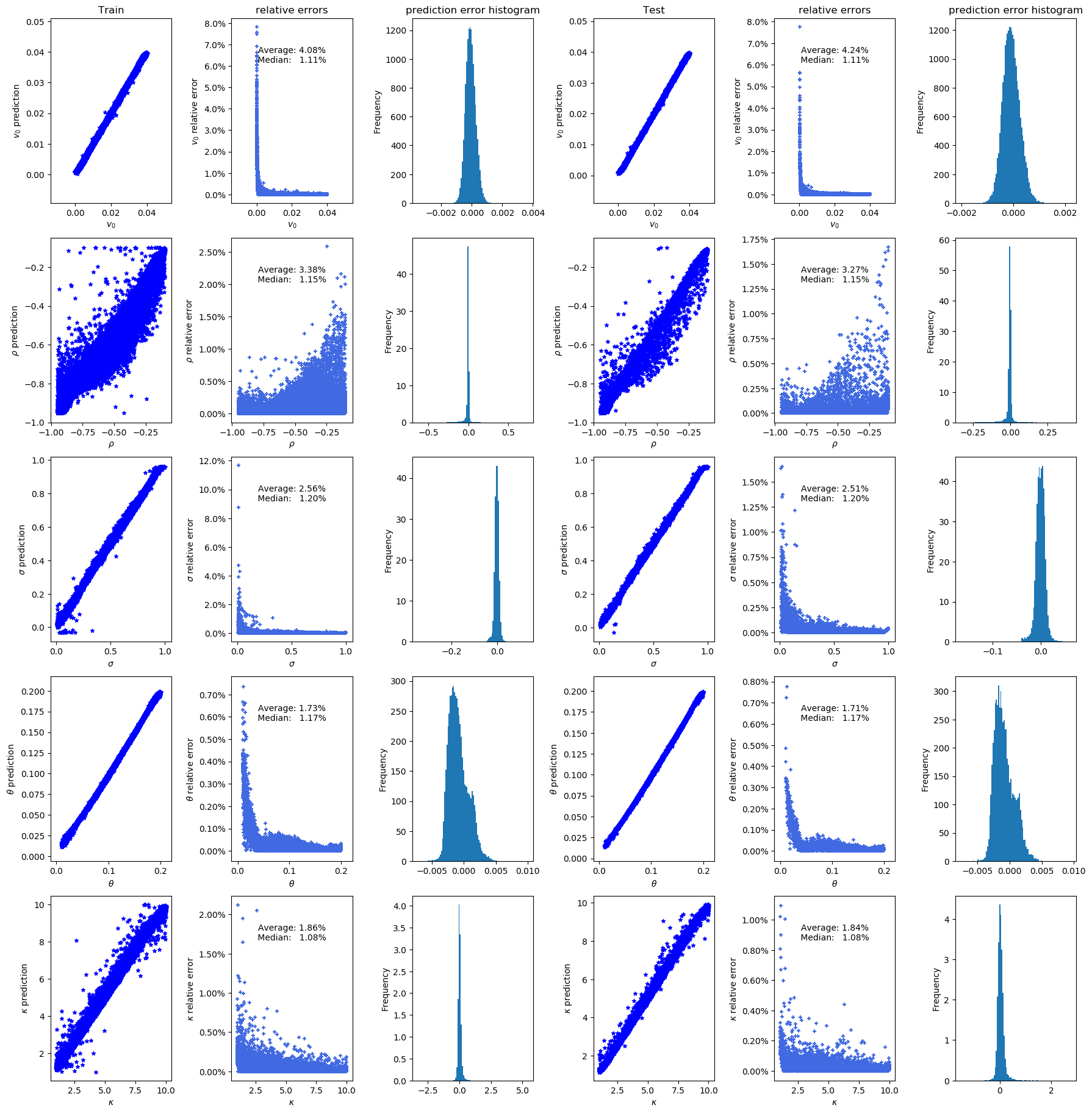}
    \caption{CNN prediction errors for large data set.}
    \label{fig:C_histlarge}
\end{figure}

\begin{figure}
    \centering
    \includegraphics[width=0.95\textwidth,keepaspectratio]{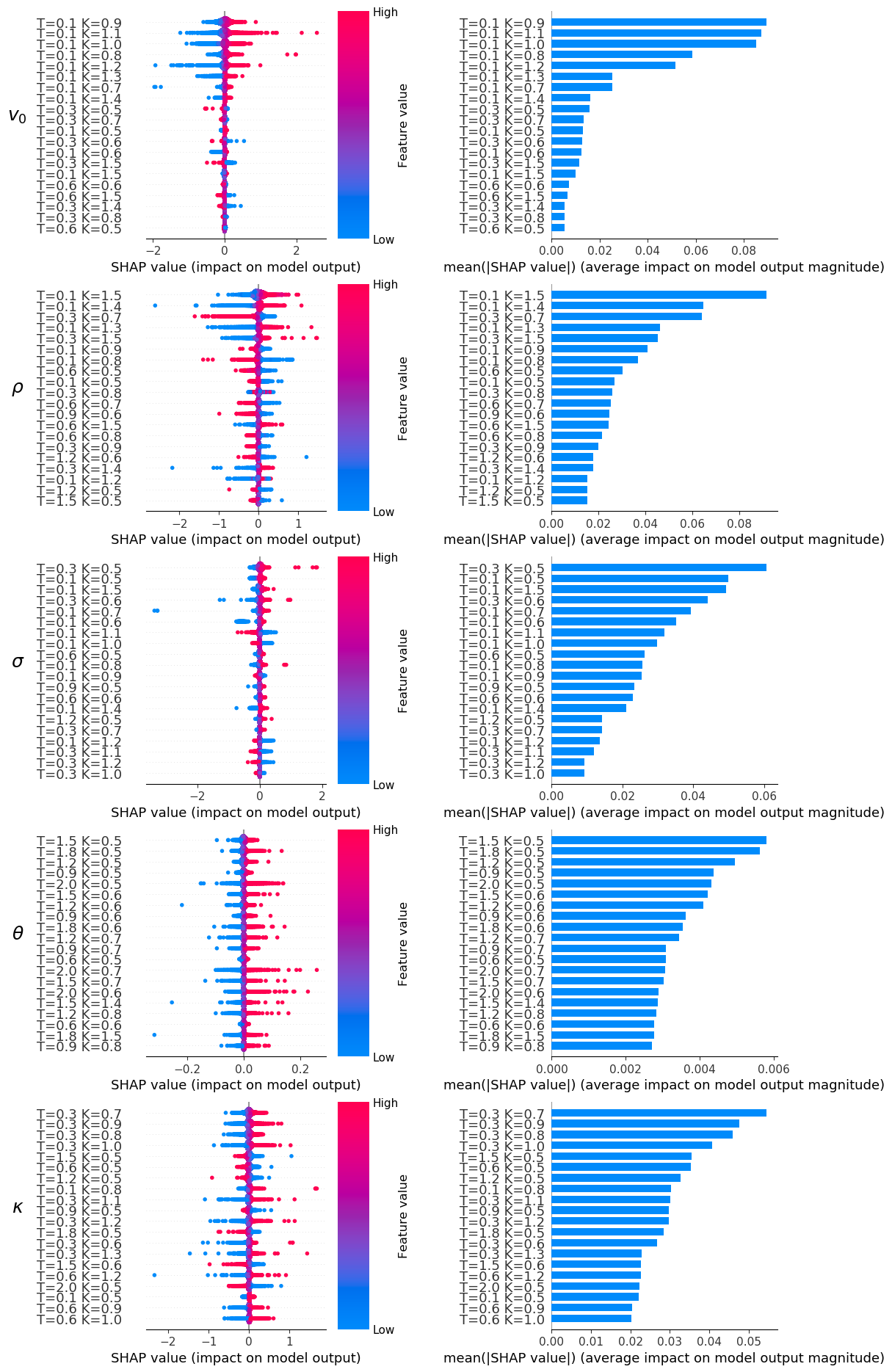}
    \caption{FCNN SHAP values and feature importance of each model parameter for large data set.}
    \label{fig:FNN_SHAP_seperatelarge}
\end{figure}

\begin{figure}
    \centering
    \includegraphics[width=0.95\textwidth,keepaspectratio]{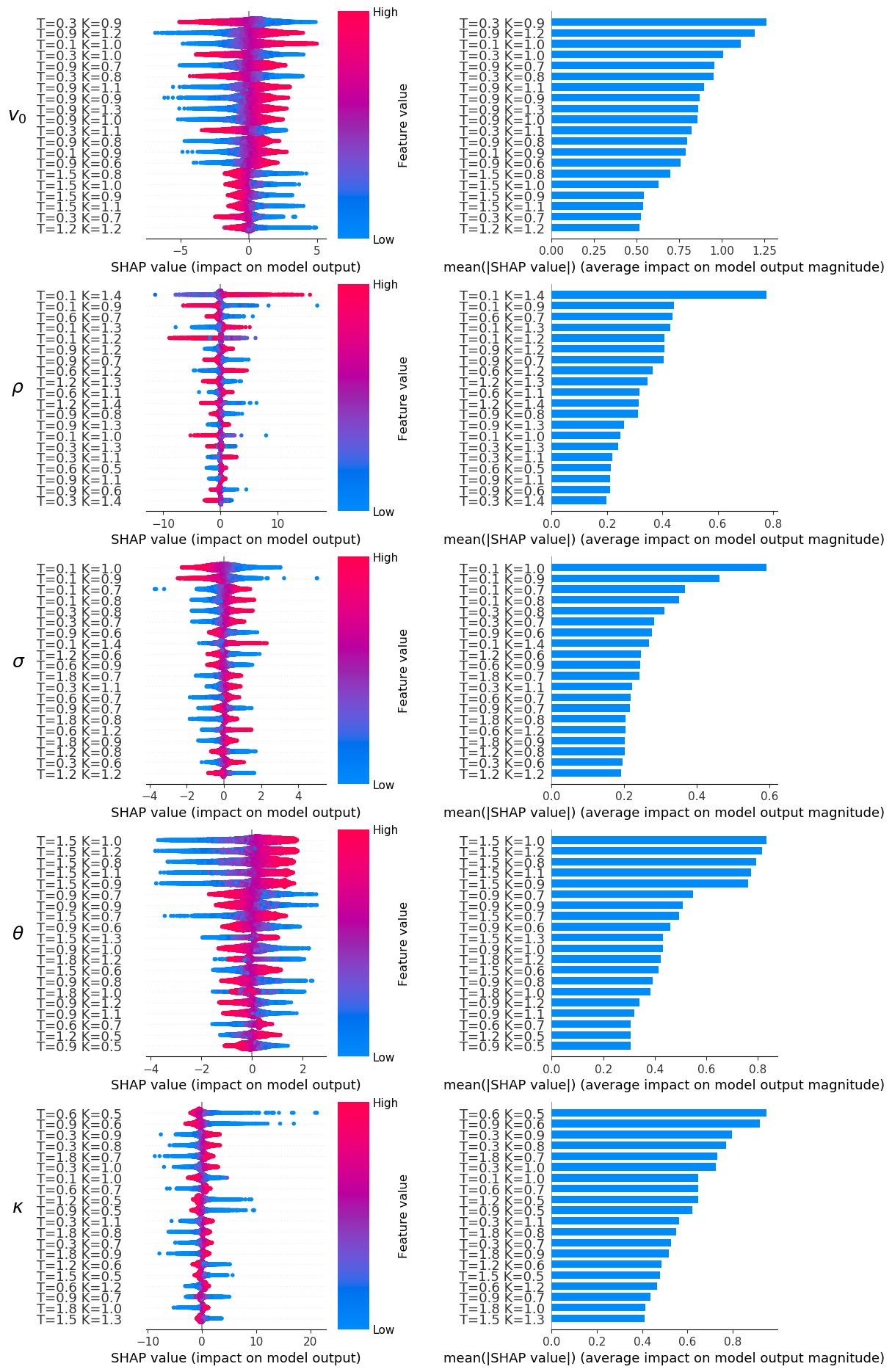}
    \caption{CNN SHAP values and feature importance of each model parameter for large data set.}
    \label{fig:CNN_SHAP_seperatelarge}
\end{figure}

\begin{figure}
	\centering
    \includegraphics[width=.50\textwidth,keepaspectratio]{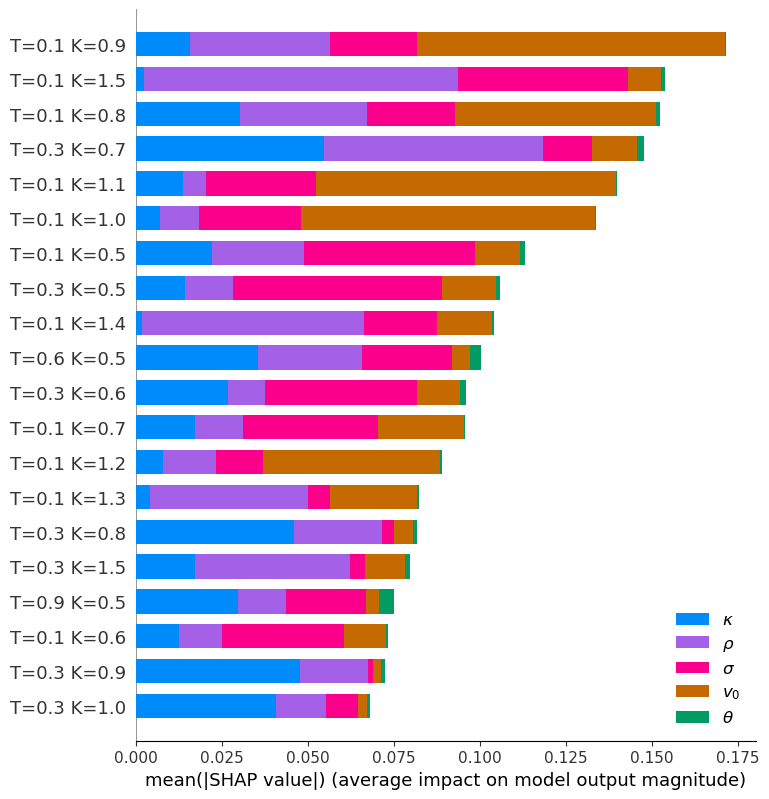}
    \hspace*{1em}
	\includegraphics[width=.40\textwidth,keepaspectratio]{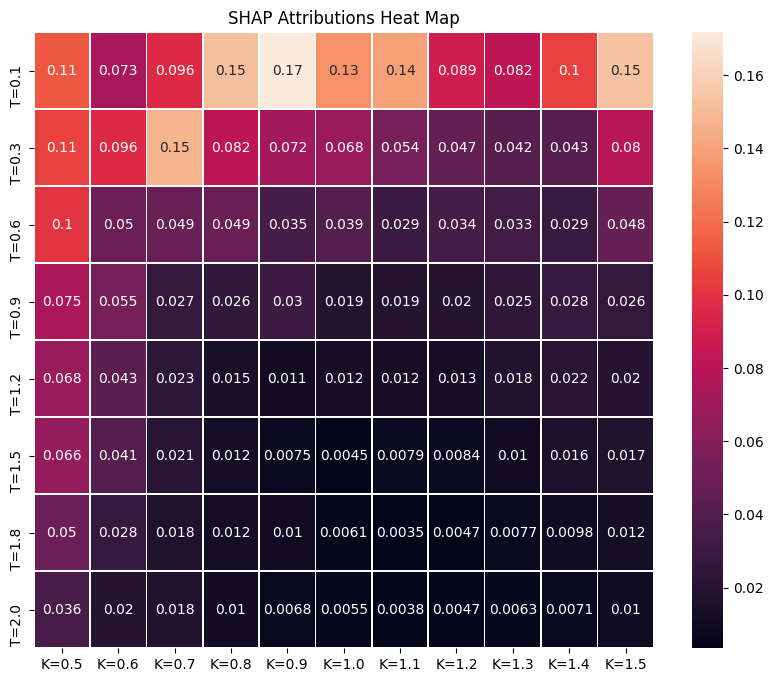}
	\caption{FCNN SHAP values overall feature importance and heat map for large data set.}
	\label{fig:SHAP_allfcnnlarge}
\end{figure}

\begin{figure}
	\centering
    \includegraphics[width=.50\textwidth,keepaspectratio]{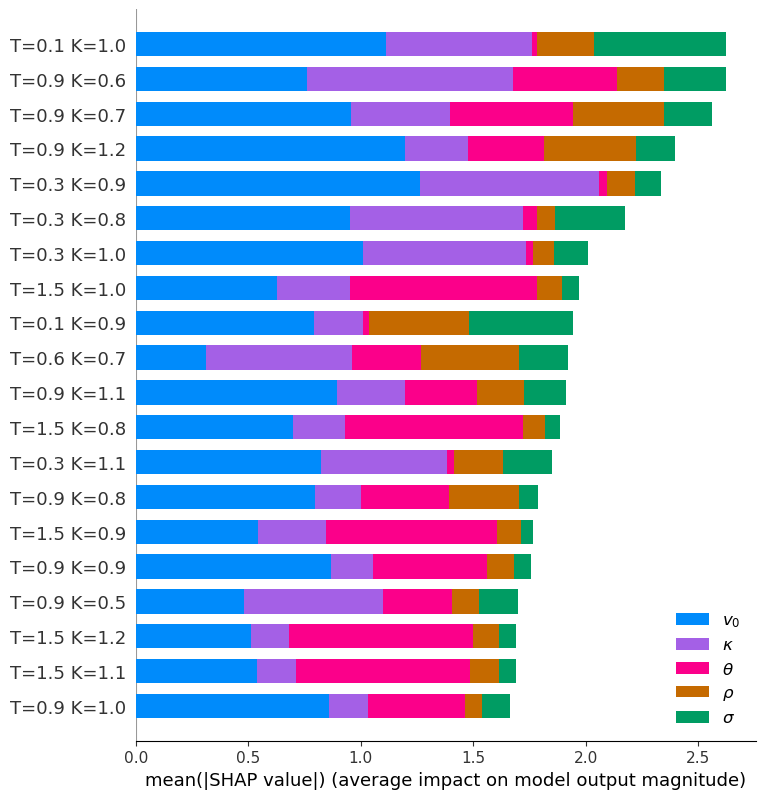}
    \hspace*{1em}
	\includegraphics[width=.40\textwidth,keepaspectratio]{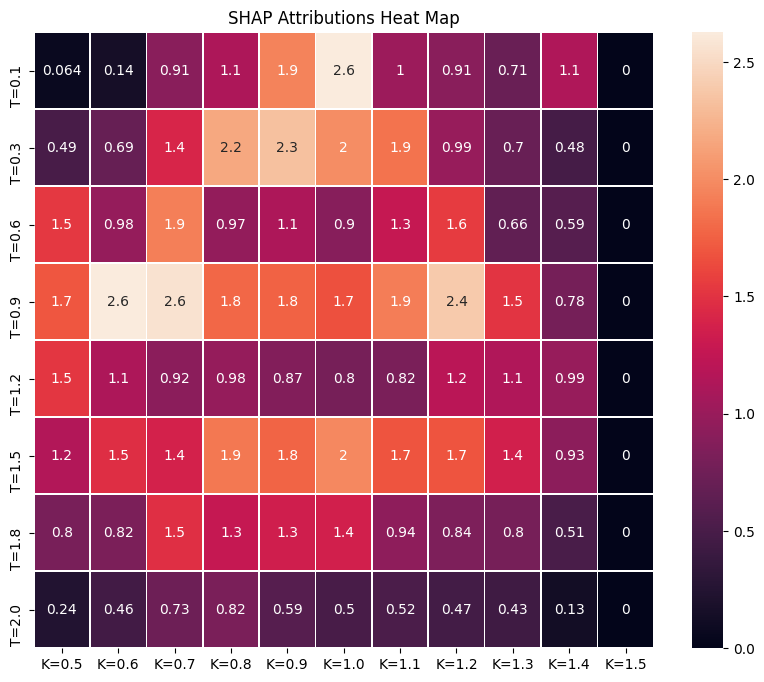}
	\caption{CNN SHAP values overall feature importance and heat map for large data set.}
	\label{fig:SHAP_allcnnlarge}
\end{figure}

We list also the results obtained using a data set of $10^{5}$ data points in Figures~\ref{fig:F_histlarge}, \ref{fig:C_histlarge}, \ref{fig:FNN_SHAP_seperatelarge}, \ref{fig:CNN_SHAP_seperatelarge}, \ref{fig:SHAP_allfcnnlarge}, and~\ref{fig:SHAP_allcnnlarge}. The calibration results improve but at a high computational cost.

\end{document}